\documentclass[12pt]{article}

\usepackage{amsmath,amssymb,amsthm,cite,enumitem,mathtools,xcolor}
\usepackage{empheq}
\newcommand*\widefbox[1]{\fbox{\hspace{2em}#1\hspace{2em}}}
\usepackage{framed}

\usepackage[margin=3cm]{geometry}
\usepackage{setspace}

 \usepackage[utf8]{inputenc} %% Unicode is now the default for Latex
\usepackage[T1]{fontenc}
\usepackage{newtxtext} % oldstyle digits in text, not math
\usepackage{bm} % bold math fonts load last
\usepackage[normalem]{ulem}  %to cross out sth (durchstreichen)

\usepackage{multicol}
\usepackage{authblk}

\title{Spatially flat FLRW spacetimes with a Big Bang \\
from matrix geometry}

\author[1]{Christian Ga\ss}
\author[1]{Harold C.~Steinacker}

\affil[1]{\small Faculty of Physics, University of Vienna \protect\\
Boltzmanngasse 5, A-1090 Vienna, Austria \protect\\
email: christian.gass@univie.ac.at, harold.steinacker@univie.ac.at}

\DeclarePairedDelimiterX{\cinner}[2]{(}{)}{#1\mkern2mu\delimsize\vert\mkern2mu\mathopen{}#2}
\DeclarePairedDelimiterX{\rinner}[2]{\langle}{\rangle}{#1\mkern2mu\delimsize\vert\mkern2mu\mathopen{}#2}

\DeclareMathOperator{\Tr}{Tr}       %% (operator) trace
\DeclareMathOperator{\realpart}{Re}

\renewcommand{\Re}{\realpart}

\newcommand{\dd}{\textup{d}}
\newcommand{\ii}{\textup{i}}
\newcommand{\ee}{\textup{e}}

\newcommand{\PdS}{\rm PdS}

\newcommand{\R}{\textup{R}}

\newcommand{\eps}{\varepsilon}      %% short for \varepsilon
\newcommand{\La}{\Lambda}           %% short for \Lambda
\newcommand{\bC}{\mathbb{C}}        %% complex numbers
\newcommand{\bN}{\mathbb{N}}        %% natural numbers
\newcommand{\bR}{{\mathbb{R}}}      %% real numbers
\newcommand{\bZ}{\mathbb{Z}}        %% integer numbers
\newcommand{\del}{\partial}         %% short for \partial
\newcommand{\aside}[1]{} %% suppress `asides' in the printed version

\newcommand{\word}[1]{\quad\text{#1}\quad} %% well-spaced word(s)
\def\wick:#1:{\,\mathopen:#1\mathclose:\,} %% Wick-ordered operator

\def\epsi^#1_#2{\eps^{#1}{}_{\!#2}} %% \eps^{ab}_c / \eps^a_{bc}
\def\epsii_#1^#2{\eps_{#1}{}^{\!#2}} %% \eps_{ab}^c / \eps_a^{bc}
\def\lLa^#1_#2{\Lambda^{#1}{}_{#2}}  %% \lLa^\mu_\nu

\def\duo<#1,#2>{\langle#1,#2\rangle} %% duality pairing <x,y>
\def\scal<#1|#2>{\langle#1\mathbin|#2\rangle} %% prod escalar <w|z>

\newcommand{\bes}{\begin{subequations}}
\newcommand{\ees}{\end{subequations}}

\theoremstyle{plain}
\numberwithin{thm}{section}         %% Number thms by section
\theoremstyle{definition}
\def\bbbone{{\mathchoice {\rm 1\mskip-4mu l} {\rm 1\mskip-4mu l}
{\rm 1\mskip-4.5mu l} {\rm 1\mskip-5mu l}}}
\def\one{\bbbone}

\def\cQ{{\mathcal Q}}

\def\cD{{\mathcal D}}
\def\cH{{\mathcal H}}

\def\cM{{\mathcal M}}

\def\cC{{\mathcal C}}

\newcommand{\beq}{\begin{equation}}
\newcommand{\eeq}{\end{equation}}

\usepackage{graphicx}
\usepackage{floatrow}
\usepackage{tikz,tikz-3dplot}
\usetikzlibrary{calc,arrows,shapes,positioning,decorations,3d,patterns}
\tdplotsetmaincoords{80}{45}
\tdplotsetrotatedcoords{-90}{180}{-90}

\def\nn{\nonumber}

\def\R{{\mathbb R}} \def\C{{\mathbb C}} \def\N{{\mathbb N}}
 \def\one{\mbox{1 \kern-.59em {\rm l}}}

\def\mso{\mathfrak{so}}
\newcommand{\End}{\mathrm{End}}
\def\hs{\mathfrak{hs}}

\def\La{\Lambda}

\tikzset{surface/.style={draw=blue!70!black, fill=blue!40!white, fill opacity=.25}}

\makeatletter
\renewcommand{\section}{\@startsection{section}{1}{\z@}%
                       {-3.5ex \@plus -1ex \@minus -.2ex}%
                       {2.3ex \@plus.2ex}%
                       {\normalfont\large\bfseries}}
\renewcommand{\subsection}{\@startsection{subsection}{2}{\z@}%
                       {-3.25ex \@plus -1ex \@minus -.2ex}%
                       {1.5ex \@plus .2ex}%
                       {\normalfont\normalsize\bfseries}}

\usepackage{parskip}

\makeatother

\numberwithin{equation}{section}

\usepackage{hyperref}
\usepackage{xcolor}
\hypersetup{
    colorlinks,
    linkcolor={blue!30!black},
    citecolor={blue!30!black},
    urlcolor={blue!30!black}
}

\begin{document}
\maketitle 
\begin{abstract} 
\noindent
We present an expanding, spatially flat ($k=0$) FLRW quantum spacetime with a Big 
Bang, considered as a background in Yang-Mills matrix models. The FLRW geometry emerges 
in the semi-classical limit as a projection from the fuzzy hyperboloid. We analyze 
the propagation of scalar fields, and demonstrate that their Feynman propagator 
resembles the Minkowski space Feynman propagator in the semi-classical regime. 
Moreover, the higher spin modes predicted by the matrix model are described explicitly. 
These results are compared to recent results on $k=-1$ FLRW quantum spacetimes with 
a Big Bounce.

\end{abstract}
{\footnotesize
\tableofcontents}

\section{Introduction}
\label{sec:Intro}
\subsection{Cosmological FLRW spacetimes from matrix geometry}
Cosmological FLRW quantum spacetimes with negative spatial curvature $k=-1$ 
and a Big Bounce have recently been described in the context of the 
IKKT matrix model \cite{BStein,SteinBB,SteinFLRW,SperlingStein}, 
see also \cite{Steinbook}. They can be described semi-classically as 
a projection of the fuzzy hyperboloid $H_n^4$. In the present paper, we 
investigate a different projection of $H_n^4$ and a corresponding matrix 
configuration, which yields a spatially flat covariant quantum spacetime. 
Its semi-classical limit is an expanding $k=0$ FLRW spacetime 
with a Big Bang. 

In the framework under consideration, a quantum geometry is defined 
by a set of hermitian matrices or operators $T^a$, which act on some 
underlying Hilbert space $\cH$. Such a "matrix configuration" may arise 
as a solution of particular matrix models, most notably the maximally 
supersymmetric IKKT model, denoted sometimes as "matrix theory". 
In this context, such a matrix configuration 
naturally defines a frame and an effective metric on a quantum space, 
by identifying $\End(\cH) \cong \cC(\cM)$ as quantized space of functions 
on some underlying symplectic space $\cM$. In particular, the matrix model 
defines a matrix d'Alembertian $\Box$, which governs the propagation of
matrix modes describing physical fields on the background.

In the first part of the paper, we discuss a matrix background defining 
a spatially flat FLRW spacetime with manifest $E(3)$ symmetry, elaborating 
on the brief exposition given in \cite{Steinbook}. This provides a new example 
of a covariant quantum spacetime in the spirit of \cite{SteinBB,SteinFLRW,SperlingStein} 
with an internal $S^2$ fiber, leading to a tower of higher-spin ($\hs$) modes 
on spacetime. In contrast to previous examples, this turns out to describe a 
cosmological quantum spacetime with a Big Bang.\footnote{For other related work 
on the origin on quantum spacetime from matrix theory see e.g. \cite{Brahma:2022hjv,Brahma:2021tkh,Asano:2024def,Nishimura:2019qal,Hirasawa:2023lpb}.}
We then consider the simplest model for a physical field on the 
new background, corresponding to free scalar field theory. The 
quantization in this model is defined by a matrix "path" integral 
over the space of hermitian matrices, which we make well-defined 
by imposing a Feynman-type $i\varepsilon$ prescription. Our aim 
is to compute the corresponding 2-point function, and to recover 
the classical propagator on Minkowski space in a local regime, 
following the lines of \cite{BStein}. This is the minimal requirement 
for any physically acceptable theory on (quantum) spacetime. The 
2-point function is a Green function of the matrix d'Alembertian 
$\square$, which arises directly from the matrix model and governs 
the path integral quantization.

\subsection{Effective scalar field propagation}
In the semi-classical regime, the matrix d'Alembertian can be related to various  differential operators that arise from the 
matrix model. One of them is the \emph{effective 
d'Alembertian} $\square_G$, which is the geometric d'Alembertian 
corresponding to the effective $k=0$ FLRW metric $G_{\mu\nu}$. The 
second one is the \emph{late time matrix d'Alembertian} $\square 
= \rho^2\square_G$, where $\rho^2$ is the \emph{dilaton}.
We explicitly find the eigenmodes of $\Box$ in the semi-classical regime, given by plane-waves in the space directions. This allows in particular to define a quantization map compatible with $E(3)$, in analogy to Weyl quantization.

We then focus on the propagator in the semi-classical regime. In general, 
the matrix d'Alembertian does not have an interpretation as a geometric 
d'Alembertian on some spacetime. It is a noteworthy coincidence that it 
\emph{is} related to a geometric d'Alembertian  in the present 
context because it can be mapped to the geometric d'Alembertian on the Poincaré 
patch of de Sitter space by a similarity transformation between two different $L^2$ spaces.

We  study the propagation of scalar fields in the semi-classical 
regime as follows. First, we derive the propagator 
of a scalar field associated with the operator $\square+m^2$. 
In contrast to the $k=-1$ case investigated in \cite{BStein}, 
we are able to derive exact formulas for this propagator in terms of 
Gegenbauer functions (or equivalently, associated Legendre functions). 
This exact solvability is no surprise due to the close relation of 
$\square$ to the geometric d'Alembertian on the Poincaré patch 
of de Sitter space (PdS). It is well-known that scalar field propagators 
on PdS can be described in terms of the mentioned special functions, 
see for example \cite{BirrellDavies,FSS13}. However, the propagators 
described in \cite{BirrellDavies,FSS13} are not the ones that 
arise from the matrix model because they either rely on restricting 
a distinguished propagator on the global patch of de Sitter space 
to the Poincaré patch or are obtained from specifying an asymptotic 
behavior of on-shell modes, i.e., from choosing a basis of the space 
of \emph{solutions} of the Klein-Gordon equation instead of a 
(generalized) basis  of $L^2(\cM)$ consisting of eigenfunctions of $\square$. 

In contrast, the path integral quantization in the matrix model formulation 
is based on the space of all off-shell modes, hence on the choice of an
orthonormal (generalized) basis of eigenfunctions of $\square$ of 
$L^2(\cM)$. This guarantees that the action and the propagator can be written
as an integral (or a sum) over these off-shell modes. 
Accordingly, we have to choose a self-adjoint realization of 
$\square$ and determine the corresponding (generalized) eigenfunctions. 
It turns out that in the present case of a spatially flat FLRW spacetime, 
we need to impose boundary conditions at temporal infinity to make $\Box$ 
self-adjoint. Details on the specific boundary conditions can be found 
in Appendix \ref{app:closed_sa_real}. 

For any possible choice of self-adjoint boundary conditions, we observe a 
reflection that causes an additional singularity of the propagator if 
$\tau+\tau'= |\vec{x}-\vec{x}'|$, where $\tau,\tau'\geq0$ are the timelike 
variables and where $\vec{x},\vec{x}'$ are the spacelike variables. This
can be interpreted as reflection at the Big Bang, which is 
the analog of the observed pre-post-Big-Bounce correlations in the $k=-1$ 
case \cite{BStein}. The propagators corresponding to self-adjoint boundary 
conditions are not the ones typically found in the literature on PdS, so we 
derive them from first principles. In fact, we do not know of any derivation 
of scalar field propagators in PdS related to the question of self-adjoint 
boundary conditions in the existing literature. However, 
there are analyses of scalar field propagators on the Poincaré 
patch of anti-de Sitter space for different boundary conditions 
\cite{DF16,DFJ18}. Moreover, self-adjointness based approaches, which 
guided our derivations, exist in many other cases 
\cite{DS18,DS19,DS22,DG24_propa,NT23,NT23_2,RumpfdS,rumpf1,V20}, 
including the global patch of de Sitter space. In the cited literature,  
there is typically no need to impose boundary conditions due to the 
essential self-adjointness of the d'Alembertian. Exceptions are a very 
general paper \cite{DS22}, where slab geometries are considered, and the 
mentioned literature on the Poincaré patch of anti-de Sitter 
space \cite{DF16,DFJ18}.

Although the propagators corresponding to $\square+m^2$ with self-adjoint 
boundary conditions are exactly described in terms of special functions, 
they are expected to provide a good description of the
matrix model only in the semi-classical regime. We therefore 
 determine their asymptotic form in this regime, which corresponds  
to late times $\tau,\tau'\gg1$ and large mass parameter $m\gg1$. The latter requirement
arises from the fact that $m$ is related to the physical mass via the dilaton.
%meaning that $m$ should grow with time. 
Moreover, we must assume that the 
time difference $|\tau-\tau'|$ in the local patch under consideration is small 
compared to $\tau$ and $\tau'$. The asymptotic form of the propagators in this 
regime turns out to be comparably simple.

 The relevance of a family of self-adjoint realizations of $\square$ 
 corresponding to different boundary conditions is somewhat unexpected 
 from the matrix model point of view. 
 It stems from the fact that to define the propagator as a mode integral, 
 we must specify the global properties (i.e., boundary conditions at 
 temporal infinity) of the space of functions on which the semi-classical 
 d'Alembertian acts. However, the semi-classical approximation is only valid 
 in the % semi-classical 
 regime described above. We argue in Subsection  
 \ref{ssc:semi_classical} that the dependence of the propagator on the 
 boundary conditions 
is  suppressed 
 in this semi-classical regime, so that 
 the observed "degeneracy" of the semi-classical theory is of minor importance.

\section{Covariant \texorpdfstring{\boldmath{$k=0$}}{} cosmological quantum spacetime}
We consider the framework of quantum spaces viewed as quantized symplectic spaces 
$\cM$, 
defined through hermitian operators or matrices acting on a (separable) Hilbert space 
$\cH$. The space of ("nice") operators in $\End(\cH)$ is then interpreted in terms of 
quantized functions on the underlying manifold $\cM$, related by some
 quantization map 
\begin{align}
\label{Q-general}
  \cQ: \quad \cC(\cM) &\to \End(\cH), \ \nn \\
  \phi(\cdot) &\mapsto \Phi.
\end{align}
In the semi-classical (infrared, IR) regime, the map $\cQ$ should be invertible 
and satisfy 
\begin{align}
\cQ(\phi\psi) \sim \cQ(\phi) \cQ(\psi) \word{and} 
i\cQ(\{\phi,\psi\}) \sim [\cQ(\phi), \cQ(\psi)],
\end{align}
where $\sim$ indicates approximate equality in the semi-classical regime.
Quantization maps with these properties are not unique, but preferred maps can 
often be found if one requires that $\cQ$ be compatible with some symmetry group, 
notably for quantized coadjoint orbits such as  fuzzy $\C P^n$ or the fuzzy 
sphere \cite{Steinbook}. 
Another way to describe quantization maps is via (quasi-)coherent states 
as $\Phi = \int \Omega \phi(x)|x\rangle\langle x|$ where $\Omega$ is the 
symplectic volume form on $\cM$; see e.g. \cite{Steinacker:2019fcb,Steinacker:2020nva,Steinbook} 
for an introduction to this framework.

In practice, defining $\cQ$ amounts to choosing suitable bases 
${\bf\Upsilon}_\La$ of $\End(\cH)$ and $\Upsilon_\La$ of $\cC(\cM)$, 
so that 
\begin{align}
\label{Q-def-general}
\cQ(\Upsilon_\La) := {\bf\Upsilon}_\La
\end{align}
respects inner products on $\cC(\cM)$ and $\End(\cH)$ 
and be compatible with pertinent symmetries. 
Hence $\cQ$ maps functions on $\cM$ to operators,  
\begin{align}
 \cC(\cM) \ni \ \phi = \sum_{\Lambda} \phi_{\La} \Upsilon_\La
 \quad \stackrel{\cQ}{\mapsto} \quad \Phi = \sum_\Lambda \phi_{\La} {\bf\Upsilon}_\La
  \quad \in  \End(\cH)
\end{align}
(symbolically; the basis need not be discrete). 

For the covariant quantum spacetimes under consideration, the symplectic space $\cM$ 
exhibits a bundle structure. Specifically, it takes the form of an $S^2$-bundle over 
the spacetime manifold $\cM^{1,3}$, indicating that the model naturally accommodates 
higher spin modes \cite{SperlingStein}. In the present case of a cosmological $k=0$ 
spacetime, these higher spin modes are particularly simple, see Subsection \ref{ssc:hs}.
The internal $S^2$ structure does not contribute to the scalar field propagation described 
in Subsections \ref{ssc:matrix_scalar} and \ref{ssc:effective_G} and Section \ref{sec:prop}, 
but it is relevant for the general case. 

\subsection{Matrix background for the covariant \texorpdfstring{\boldmath{$k=0$}}{} cosmological quantum spacetime}
Specifically, we consider the covariant $k=0$ cosmological quantum spacetime $\cM^{1,3}$ briefly introduced in \cite{Steinbook}.
This is defined in terms of a unitary representation $\cH = \cH_n$
of $\mso(4,2)$ in the minimal discrete ("doubleton") series \cite{FernandoGunaydin} for $n\in \N$ as follows:
let $M^{ab}$ be the generators of $\mso(4,2)$, which satisfy 
\begin{align}
  [M^{ab},M^{cd}] &=\ii (\eta^{ac}M^{bd} - \eta^{ad}M^{bc} - \eta^{bc}M^{ad} + \eta^{bd}M^{ac}) \ 
 \label{M-M-relations}
\end{align}
for $a = 0,\ldots ,5$.
Indices will be raised and lowered with  $\eta_{ab}= {\rm diag}(-1,1,1,1,1,-1)$.
Then define the following hermitian generators
\begin{subequations}
\begin{align}
 X^a &:= M^{a c} \alpha_c  = r M^{a5}, 
 \qquad  \alpha_c = r(0,0,0,0,0,1)\label{X-def} \\
 T^\mu &= M^{\mu c} \beta_c
 = \frac 1R (M^{\mu 0} + M^{\mu 4}), \qquad  \beta_c = R^{-1}(1,0,0,0,1,0)
 \label{T-def}
\end{align}
\end{subequations}
for $a = 0,\ldots ,4 $ and $\mu=0,...,3$.
Here $r$ is a scale parameter of dimension length, and $R$ is given by 
\begin{align}
X_a X^a &= - R^2 \one, \qquad R^2 = \frac{r^2}{4}(n^2-4).
\label{XX-R-fuzzy-H4}
\end{align}
Using $\beta_a \beta^a = 0 = \alpha_a \beta^a$, it is easy to 
see that the above generators satisfy the commutation relations
\begin{subequations}
\begin{align}
 [X^\mu,X^\nu] &= -\ii r^2M^{\mu\nu}
 % (?) = -\frac{4}{g^2}\int\ d^4x\rho_\cM\, \tilde\Box B_{\dot\alpha}\,T^{\dot\alpha},
  \label{X-X-CR} 
  \\ 
 [T^\mu,T^\nu] &= \ii\big(  - T^\mu \beta^\nu + \beta^\mu T^\nu \big),
 \label{T-T-CR}
 \\ 
 [X^\mu,T^\nu] &=  \ii\big( \beta^\mu X^\nu -\frac 1R\eta^{\mu\nu}\,(X^0 + X^4)\big),
 \label{X-P-CR-general}
 \\ \label{X4-T-CR}
 [X^4,T^\mu] &= \frac{\ii}{R} X^\mu .
 \end{align}
 \end{subequations}
  Explicitly, the $T^\mu, \ \mu=0,...,3$ satisfy the relations\footnote{The 
  commutation relations of the $T^\mu$ are sometimes denoted as $\kappa$ 
  Minkowski space. It is essential here that they arise as a sub-algebra 
  of a larger algebra.}
 \begin{align}
 \boxed{\ 
 [T^i,T^j] = 0 , \qquad  [T^0,T^i] = -\frac \ii R T^i \ 
 }
 \label{TT-CR}
 \end{align}
as well as 
\begin{align}
   [T^i,X^j] &= \frac{\ii}{R}\delta^{ij}(X^0+X^4), \nn\\
    [T^i,X^0] &=  \frac {\ii}{ R} X^i, \nn\\
    [T^0,X^j] &=0, \nn\\
     [T^0,X^0] &=  -\frac{\ii}{R} X^4 .
\end{align}
Here, $X^4$ is a function of $X^\mu$ due to the constraint \eqref{XX-R-fuzzy-H4}.
These relations simplify for 
\begin{align}
    Y^i = X^i, \qquad Y^0 = R X^a \beta_a = X^0 + X^4 \ ,
\end{align}
which satisfy the constraints 
\begin{align}
T_i T^i &= \frac 1{R^2 r^2} Y_0^2 , \nn\\
T_i Y^i + Y^i T_i &= T^0 Y^0 + Y^0 T^0 
\label{T-Y-constraints}
\end{align}
 implied by \eqref{MM-contraction-42}. As discussed below, these constraints 
 provide the basis for interpreting 
$\End(\cH_n)$ in terms of quantized functions on an $S^2$ 
bundle over spacetime, in complete analogy to the results 
in \cite{SperlingStein} for $k=-1$.
The commutation relations involving $Y^\mu$ are 
 \begin{subequations}
 \label{T-Y-CR}
 \begin{empheq}[box=\widefbox]{align}
    [T^\mu, Y^\nu] &= \ii\frac{Y^0}{R}\eta^{\mu\nu},
    \label{T-Y-flat-k=0}
\\  \label{Y0-Yi-CR}
[Y^0,Y^i] &= \ii r^2 R T^i,
\\ \label{Yi-Yj-CR}
 [Y^i,Y^j] 
&= -\ii r^2 R(Y^0)^{-1} (T^i Y^j - T^j Y^i) .  \phantom{\Big(}
\end{empheq}
\end{subequations}
The last formula  is specific for the $n=0$ case, while for $n>0$ 
there is an extra term with $\varepsilon^{\mu\nu\rho\sigma}$ 
\cite{SperlingStein}. As a check, it is easy to verify that the 
rhs commutes with $Y^0$ and $T^0$. This agrees with the commutation 
relations of the minimal $k=-1$ covariant quantum spacetime in the 
local regime \cite{HKS}. The commutation relations 
\eqref{Y0-Yi-CR} follow immediately from \eqref{X-X-CR}. We derive 
the commutation relations \eqref{Yi-Yj-CR} in Appendix 
\ref{app:group-id}.

A comment on $n$ is in order. Even though \eqref{XX-R-fuzzy-H4} seems to 
require $n\geq 3$, it turns out \cite{MantaStein25} that the semi-classical 
geometry is essentially independent of $n$ as long as $r^{-1} x^0 \gg 1$; 
in particular, the positive sign in the rhs of \eqref{XX-R-fuzzy-H4} for 
small $n$ is a quantum effect and misleading. One can therefore restrict 
to the minimal case $n=0$ for simplicity, where the following semi-classical 
considerations hold at sufficiently late times $r^{-1} Y^0 \gg 1$. 
Nevertheless we will consider generic $n$ in this paper.

\paragraph{Quantization map.}
A physical understanding of the noncommutative algebra under 
consideration is provided by a quantization map \eqref{Q-general},
which allows to interpret $\End(\cH)$ as quantized algebra of functions 
on an underlying symplectic manifold $\cM$, at least in some semi-classical 
(IR) regime. In the present case, there are two natural 
options to define $\cQ$: 
\begin{itemize}
\item[(1)] in terms of suitable (quasi-) coherent states, or 
\item[(2)] by requiring compatibility with $E(3)$ in analogy to Weyl quantization. 
\end{itemize} 
The first option is discussed in detail in \cite{MantaStein25}, and allows to 
conclude that the semi-classical (symplectic) manifold underlying $\End(\cH)$ for 
a doubleton irrep $\cH = \cH_n$ of $SO(4,2)$ is given by $\cM \cong \C P^{1,2}$, 
known as twistor space. This is also confirmed by the second option, which amounts 
to choosing a suitable basis ${\bf\Upsilon}_\La$ of $\End(\cH_n)$ and $\Upsilon_\La$ 
of $\cC(\cM)$; and defining $\cQ$ via \eqref{Q-def-general} so that it is 
compatible with $E(3)$ and respects inner products. 
The pertinent bases will be specified 
in the following in terms of irreps of $E(3)$, i.e., plane waves on $\R^3$, and 
an appropriate basis of functions generated by the time-like generators $Y^0 \sim y^0$.

\paragraph{Symplectic geometry and bundle structure.}

Given a quantization map $\cQ$, we can describe the symplectic space $\cM$ 
underlying $\End(\cH)$ more explicitly as an $S^2$ bundle over spacetime.
The bundle structure is exhibited by the semi-classical generators 
$y^\mu = \cQ^{-1}(Y^\mu)$ and $t^i = \cQ^{-1}(T^i)$, which satisfy the 
following constraints obtained from \eqref{T-Y-constraints} in the 
semi-classical regime:
\begin{align}
t_i t^i &= \frac 1{R^2 r^2} y_0^2 , \qquad
t_i y^i  = t^0 y^0 .
\label{t-y-constraints}
\end{align}
Since the spectrum of $Y^0 \sim y^0$ is positive on  $\cH_n$, the symplectic 
space $\cM$ can be identified with $\cM^{1,3}\times S^2$, where $y^\mu$  serve 
as coordinates on the base space(time) $\cM^{1,3} \cong \R_+ \times \R^3 $, 
and $t^i$ parametrize the $S^2$ fiber. This bundle space is equipped with 
non-degenerate Poisson brackets arising from the commutation relations 
\eqref{T-Y-CR} in the semi-classical regime.

\subsection{Matrix model formulation of scalar fields}
\label{ssc:matrix_scalar}
To motivate the following considerations, we consider a matrix model 
describing a (free, noncommutative) scalar field $\Phi$ on the quantum space 
considered above. This is described by the action
\begin{align}
\label{eq:matrix_action}
 S[\Phi] = \Tr\big( [T^\mu, \Phi^\dagger] [T_\mu,\Phi] - m^2 \Phi^\dagger \Phi \big)
 =-\Tr\big(\Phi^\dagger (\Box + m^2)\Phi \big),
\end{align}
which can be seen as a toy model for the  bosonic part of the 
action of the IKKT model with a static instead of dynamical
background $T^\mu$ in the regime where the matrices describing 
the extra dimensions, represented by several scalars, couple weakly. 
In \eqref{eq:matrix_action}, $\Box$ denotes the "Matrix d'Alembertian" 
\begin{align}
\label{matrix-dalembertian}
\Box :=  [T^\mu,[T_\mu,\cdot]]: \quad \End(\cH) \to \End(\cH),
\end{align} 
acting on the space of operators $\End(\cH)$, which is interpreted as space 
of quantized functions. Here, the matrix indices are contracted with $\eta_{\mu\nu}$.

In the semi-classical regime, the action reduces to
\begin{align}
\label{eq:semicl_action}
 S[\Phi] \sim -\int_\cM \Omega \Big( \{t^\mu,\phi^\ast\} \{t_\mu,\phi\} 
        +m^2 \phi^\ast \phi \Big)
 = -\int_\cM \Omega \phi^*(\Box + m^2)\phi,
\end{align}
where $\Omega$ is the symplectic volume form on $\cM$ \cite{Steinbook}.
We have made a slight abuse of notation, denoting also the 
semi-classical d'Alembertian by $\Box$, 
\begin{align}
\label{eff-dAlembertian}
\Box := -\{ t^\mu,\{t_\mu,\cdot\}\} =  \rho^2 \Box_G ,
\end{align}
acting on a suitable space of functions on $\cM$, where $\rho^2$ is 
the dilaton field and 
\begin{align}
\Box_G = - \frac{1}{\sqrt{-G}} \del_\mu \sqrt{-G} G^{\mu\nu} \del_\nu
\end{align}
is the \emph{effective d'Alembertian}.  Note that with our 
sign conventions, the Minkowski metric $\eta_{\mu\nu}$ has a minus sign 
in the time component, while the d'Alembertian has a plus sign 
in the time component. We describe the appropriate function space, the 
dilaton and the effective metric and d'Alembertian in the following 
subsections \ref{ssc:effective_G} and \ref{ssc:eigenfunctions}. 

The scalar field $\phi$ only depends on the spacetime coordinates 
$y^\mu$ but not on the coordinates parametrizing the internal $S^2$. 
Therefore, the integral over the internal $S^2$ is trivial and the 
semi-classical action \eqref{eq:semicl_action} for a scalar field 
reduces to an integral over four-dimensional spacetime $\cM^{1,3}$ 
with respect to the invariant volume form $\Omega^{1,3}$ described 
in Appendix \ref{sec:invar-vol-form}, 
\begin{align}
\label{eq:semicl_action_scalar}
S[\Phi] \sim  -\int_{\cM^{1,3}} \Omega^{1,3} \phi^*(\Box + m^2)\phi.
\end{align}

The basic observables of interest are $n$-point correlators of 
matrices or operators under the matrix 
"path" integral. To make this explicit, consider some basis 
$\{{\bf\Upsilon}_\La\}$ of (a suitable subspace of) $\End(\cH)$. 
Then the 2-point correlation functions are defined as 
\begin{align}
\label{matrix-integral}
\langle \Phi_{\Lambda} \Phi_{\La'}\rangle 
&:= \frac 1Z\,\int D\Phi\, \Phi_{\La} \Phi_{\La'} 
e^{i S_\varepsilon[\Phi]}  
\end{align}
 $D\Phi = \Pi_\La d\Phi_{\La}$ is the integral over all  modes, 
and the  $\varepsilon$ indicates a suitable regularization for 
the oscillatory integral. This is achieved by an $i\varepsilon$ 
deformation of the kinetic action, through a "matrix Wick rotation" 
\begin{align}\label{eq:eps_wiggle}
% T^0 \to T^0 e^{i\varepsilon}, \qquad 
\Box \to \Box_\varepsilon := -  e^{2\ii\varepsilon} [T^0,[T^0,.]] 
 + e^{-2\ii\varepsilon} \sum\limits_i [T^i,[T^i,.]]
\end{align}
for small $\varepsilon \searrow 0$. Then the Gaussian integral 
over the space of hermitian matrices \eqref{matrix-integral} is 
well-defined at least for finite-dimensional matrices, and we 
will adopt this also for infinite-dimensional matrices.

 The physical significance
of the above correlation functions can be seen using the $E(3)$ - 
equivariant quantization map \eqref{Q-def-general}.
We can then associate the above correlators to 2-point 
function on $\cM$ as follows\footnote{The hat on the lhs indicates 
that this "2-point function" lives in the dual of $\cC(\cM)^{\otimes 2}$.} 
\begin{align}
\label{2-point-function}
D(x,y) := \langle \hat\phi(x) \hat\phi(y)\rangle 
&:=  \sum_{\La,\La'}\Upsilon_\La(x) \Upsilon_{\La'}(y) 
  \langle \Phi_{\La} \Phi_{\La'}\rangle \ .
\end{align}
This will be interpreted as a 2-point function in some noncommutative deformation of (free, in the present case) quantum field theory on $\cM$.
These correlation functions can be computed as usual in terms of Gaussian integrals with respect to the sesquilinear form 
\begin{align}
 S_{\La \La',\eps} 
% = \langle {\bf\Upsilon}_\La {\bf\Upsilon}_{\La'}\rangle 
 :=  -\Tr\big({\bf\Upsilon}_\La ^\dagger (\Box_\eps + m^2){\bf\Upsilon}_{\La'} \big) = S_{\La' \La,\eps}^*
\end{align}
% or the bilinear form 
% \begin{align}
%  S_{\La \La'} = \langle {\bf\Upsilon}_\La,{\bf\Upsilon}_{\La'}\rangle :=  -\Tr\big({\bf\Upsilon}_\La  (\Box + m^2){\bf\Upsilon}_{\La'} \big) 
% \end{align}
Then the 2-point function \eqref{matrix-integral} is 
computed by a Gaussian integral:
\begin{align}
\langle \Phi_{\Lambda} \Phi_{\La'}\rangle 
&= \frac 1Z\,\int D\Phi\, \Phi_{\La} \Phi_{\La'} 
e^{i S_\varepsilon[\Phi]}  
 = (S_\varepsilon^{-1})_{\La\La'} \ .
\end{align}
This incorporates Feynman's $i\varepsilon$ prescription for the propagator,
as familiar from the path-integral formulation of quantum field theory. 
By construction, the 2-point function \eqref{2-point-function} 
then satisfies
\begin{align}
(\Box_{x} + m^2)D(x,y) = \delta^{(4)}(x,y)
=: y^0 \delta^{(4)}(x-y)
\end{align}
in the semi-classical regime,
where $\delta^{(4)}(x-y)$ is the flat Dirac delta.

Some comments are in order. Since the quantization map $\cQ$ should only be 
trusted in the semi-classical IR regime, the above 2-point functions (and 
generalizations thereof) should be expected to correspond to "ordinary" 
2-point functions only in the semi-classical regime. The UV regime of such a 
noncommutative field theory is in general very different from ordinary field 
theory, and should be considered with great caution.

\subsection{Effective FLRW metric and d'Alembertian}
\label{ssc:effective_G}
In the semi-classical limit $Y^\mu\sim y^\mu=(y^0,x^1,x^2,x^3) =:(\tau,\vec{x})$,
the commutation relations \eqref{T-Y-flat-k=0}
% \begin{align}
%  [T^\mu,Y^\nu] = \ii \frac{Y^0}{R} \eta^{\mu\nu}
% \end{align}
define a frame $E^{\alpha\mu} := \{t^\alpha,y^\mu\} 
= \frac{\tau}{R} \eta^{\alpha\mu}$ on spacetime $\cM^{1,3}$. 
 Inspecting the action \eqref{eq:semicl_action} 
 and requiring that it has standard geometric form expressed 
 in terms of a metric $G_{\mu\nu}$, one can easily extract the 
 form of $G_{\mu\nu}$ as\footnote{Confer also \cite{SperlingStein} 
 or Sections 9.1, 10.1 and 10.2 of \cite{Steinbook}.} 
\begin{align}
G_{\mu\nu} := \rho^2 \gamma_{\mu\nu}, 
\qquad \rho^2 = \rho_M \sqrt{|\gamma^{\mu\nu}|},
\end{align}
where
$\gamma^{\mu\nu} := \eta_{\alpha\beta} E^{\alpha\mu} E^{\beta\nu}$.
Inserting the concrete form of the frame and the density $\rho_M$, 
we obtain
\begin{align}
 G_{\mu\nu} = \tau R \, \eta_{\mu\nu}, \qquad 
  \rho^2 = \frac{{\tau}^3}{R} \ .
\end{align}
The line element corresponding to $G_{\mu\nu}$ is
\begin{align}
 \dd s_G^2 = \tau R \, \big(-\dd {\tau}^2 + \dd\vec{x}^2\big),
\end{align}
and the associated d'Alembertian reads
\begin{align}
 \square_G = - \frac{1}{\sqrt{-G}} \del_\mu \sqrt{-G} G^{\mu\nu} \del_\nu
 = \frac{1}{\tau R} \Big( \del_{\tau}^2 + \frac{1}{\tau}  
 \del_{\tau} - \Delta_3 \Big).
\end{align}
For $t=\frac{2}{3} {\tau}^{\frac32}$, we have $\dd t = {\tau}^{\frac12} \dd \tau$. Using $t$ as a variable, the line element is
\begin{align}
 \dd s_G^2 = R \big( -\dd t^2 + a(t)^2 \dd\vec{x}^2\big) \word{with}
 a(t) = \big(\tfrac32 t\big)^{\frac13},
\end{align}
which is the line element of a spatially flat expanding FLRW metric 
with a Big Bang at $t=0$.

For the matrix model action to be well-defined, it is natural to require
that all admissible "functions" $\Phi \sim \phi(x)$ are elements in
$L^2(\cM^{1,3},\Omega^{1,3})$, where $\Omega^{1,3}=\rho_M \dd\tau\dd\vec{x}$ 
is the invariant volume form \eqref{eq:symplectic_form}. This is the semi-classical 
limit of Hilbert-Schmidt operators on $\cH$. We also note that $\Box$ 
is then a hermitian operator on $C_c^\infty(\cM^{1,3})$ in the sense of 
$L^2(\cM^{1,3},\Omega^{1,3})$. By \eqref{eq:semicl_action_scalar}, the 
semic-classical action for scalar fields arising from the matrix action 
\eqref{eq:matrix_action} can be written using the formula 
\eqref{eq:symplectic_form} for the  invariant volume form on $\cM^{1,3}$,
\begin{align}
\label{eq:matrix_action-explicit}
 S[\Phi] 
  \sim -\int_0^\infty \frac{\dd \tau}{\tau}  \int \dd^{3}\vec{x}\;
 \phi^\ast(\tau,\vec{x})  (\square+m^2) \phi(\tau,\vec{x}),
\end{align}
where
\begin{align}
\square+m^2
= \rho^2 \square_G + m^2 \ .
\end{align}

The matrix d'Alembertian then reads
\begin{align}
\square
=  R^{-2} \Big( {\tau}^2 \del_{\tau}^2 + \tau  \del_{\tau} - {\tau}^2 \Delta_3 \Big).
\end{align}
On the semi-classical level, the $\ii\eps$ regularization 
\eqref{eq:eps_wiggle} amounts to $\Box\to\Box-\ii\eps$, or 
equivalently $\Box+m^2$ to $\Box+m^2-\ii\eps$. We may thus 
formally absorb the $\ii\eps$ regularization into the mass. 
To see this, note that \eqref{eq:eps_wiggle} implies that 
\begin{align}
\square \to 
R^{-2} \Big( {\tau}^2 \del_{\tau}^2 + \tau  \del_{\tau} 
            - {\tau}^2 \Delta_3 \Big)
+\ii \eps R^{-2}  \Big( {\tau}^2 \del_{\tau}^2 + \tau  \del_{\tau} 
            + {\tau}^2 \Delta_3 \Big),
\end{align}
and that the spectrum of ${\tau}^2 \del_{\tau}^2 + \tau  \del_{\tau} 
+ {\tau}^2 \Delta_3$ (with a $+$-sign in front of the spatial Laplacian) 
is $]-\infty,0]$ \cite{DeLee}.

 We want to point out an interesting coincidence. Namely, in the present 
case, the matrix Klein-Gordon operator turns out to be mathematically related to the 
Klein-Gordon operator on the \emph{Poincaré patch $\PdS_4$ of de Sitter space} 
via a unitary map $L^2(\cM,\Omega)\to L^2\big(\PdS_4,\sqrt{-\tilde{G}}\big)$. Namely, 
\begin{align}
 \tau^{\frac{3}{2}}(\square+m^2) \tau^{-\frac{3}{2}}
 &=  R^{-2} \Big( {\tau}^2 \del_{\tau}^2 -2 \tau  \del_{\tau} - {\tau}^2 \Delta_3 \Big)
 +m^2 +\frac{9}{4R^2} \notag \\
 &=:\; \square_{\tilde{G}} + \tilde{m}^2,
\end{align}
where $\tilde{m}^2 := m^2 +\frac{9}{4R^2}$ and where the operator 
$\square_{\tilde{G}}$ is the geometric d'Alembertian arising from 
the metric
\begin{align}
 \tilde{G}_{\mu\nu} := \frac{R^2}{\tau^2} \eta_{\mu\nu},
\end{align}
which is the metric on the $\PdS_4$.

The propagator associated to a scalar field on $\PdS_4$ is often simply
written down in terms of a specific associated Legendre function 
${\bf P}^{-\mu}_{\nu-\frac12}(-u-\ii0)$, see for example \cite{BirrellDavies}
(or equivalently in terms of the Gegenbauer function ${\bf S}_{\mu,\nu}(-u-\ii0)$
described in Appendix \ref{app:Bessel}). Here, $u$ is the inner product 
in the ambient five-dimensional Minkowski space into which $\PdS_4$ 
can be embedded.
This propagator is obtained as a restriction of the propagator 
correspondingto the so-called \emph{Bunch-Davies state} or 
\emph{Euclidean vacuum} on the full de Sitter space, which is in turn 
obtained via analytic continuation ("Wick rotation") from the sphere.

As explained in the introduction, using this propagator is
insufficient in the matrix model formulation. The matrix action
"does not know" that the Poincaré patch is half of a larger space, 
and therefore we should derive the propagator from first principles. 
Hence we should describe the quantization map using off-shell 
modes, i.e., (generalized) eigenfunctions of $\Box$.

\subsection{Eigenfunctions of the semi-classical d'Alembertian}
\label{ssc:eigenfunctions}
To find a suitable set of orthogonal eigenfunctions 
$\Upsilon_\Lambda(\tau,\vec{x})$ of the 
operator $\square$, we make a separation ansatz  $\Upsilon_\Lambda(\tau,\vec{x})
= \psi(\tau)\xi(\vec{x})$ yields the equations 
\begin{align}
-\Delta_3 \xi(\vec{x}) &= \beta^2 \xi(\vec{x}),
\label{eq:spatial_eigeneq}\\\label{eq:temporal_eigeneq}
\big(\tau^2\del_\tau^2 + \tau\del_\tau + \beta^2 \tau^2 \big)\psi(\tau) &= \lambda^2\psi(\tau).
\end{align}
Clearly, we may choose plane waves $\xi_{\vec{k}}(\vec{x}) 
:= (2\pi)^{-3} \ee^{\ii\vec{k}\vec{x}}$ as an orthogonal basis of 
(generalized) eigenfunctions of the spatial Laplacian. Therefore, the since 
$\beta^2=\vec{k}^2\geq0$, the temporal equation 
\eqref{eq:temporal_eigeneq} becomes the Bessel equation
\begin{align}
\label{eq:temporal_eigeneq1}
\big(\tau^2\del_\tau^2 + \tau\del_\tau + \vec{k}^2 \tau^2 - \lambda^2 \big)\psi(\tau) &= 0.
\end{align}
Although later we are only interested in $\lambda^2\in\bR$, let 
us for now assume $\lambda^2\in\bC$, or rather $\Re(\lambda)>0$ 
or $\lambda=\ii\mu$ with $\mu\geq0$. Then the space of solutions 
of \eqref{eq:temporal_eigeneq1} for fixed $\lambda$ is spanned by 
the \emph{Hankel functions} $H^{(+)}_\lambda(|\vec{k}| y)$ and 
$H^{(-)}_\lambda(|\vec{k}| y)$. See Appendix \ref{app:Bessel} 
for details on Bessel and Hankel functions. 

Writing from now on $k:=|\vec{k}|$, the general eigenfunction 
of the timelike equation is 
\begin{align}
\psi_{\lambda,k}(\tau) = a_\lambda H^{(+)}_\lambda(k \tau)
+ b_\lambda H^{(-)}_\lambda(k \tau).
\end{align}
We may use the connection formula \eqref{eq:connect_HJ} to express 
the Hankel function in terms of Bessel functions with a definite 
behavior at $\tau=0$ to find that 
\begin{align}\label{eq:temporal_eifenfct_0}
\psi_{\lambda,k}(\tau) \sim \frac{\ii}{\sin(\pi\lambda)} 
\Big( \frac{a_\lambda \ee^{-\ii\pi\lambda} -b_\lambda \ee^{\ii\pi\lambda} }{\Gamma(1+\lambda)} \, \big(\tfrac{k\tau}{2}\big)^\lambda
- \frac{a_\lambda - b_\lambda}{\Gamma(1-\lambda)} \big(\tfrac{k\tau}{2}\big)^{-\lambda}
\Big) \word{as} \tau\downarrow0.
\end{align}
The eigenfunctions $\psi_{\lambda,k}(\tau)$ should be square integrable 
(in the sense of wave packets) on $\bR_+$ with respect to the 
measure $\frac{\dd \tau}{\tau}$ induced by the symplectic density. If 
$\lambda=\ii\mu$ with $\mu\geq0$, then the behavior 
\eqref{eq:temporal_eifenfct_0} does not give any constraints on 
$a_\lambda$ and $b_\lambda$. However, if $\Re(\lambda)>0$, then 
\eqref{eq:temporal_eifenfct_0} implies $a_\lambda=b_\lambda$. 
Insisting on real eigenvalues, we thus obtain two types 
of eigenfunctions of \eqref{eq:temporal_eigeneq1} with appropriate 
behavior near $\tau=0$: 
\begin{align}
\psi_{\ii\mu, k}(\tau) &= a_{\ii\mu} H^{(+)}_{\ii\mu}(k \tau)
+ b_{\ii\mu} H^{(-)}_{\ii\mu}(k \tau), &&\quad \mu\geq0; 
\notag \\ \label{eq:eigenfunctions_lambda_imu}
\psi_{\lambda, k}(\tau) &= c_\lambda J_\lambda(k\tau), &&\quad \lambda>0.
\end{align}
All of these eigenfunctions are square integrable as $\tau\to\infty$, 
which can be read off the asymptotic behavior \eqref{eq:asymp_H} 
and \eqref{eq:asymp_J}.

However, these are far too many eigenfunctions to form a 
generalized orthonormal 
basis of $L^2\big(\bR_+,\tfrac{\dd \tau }{\tau}\big)$. Many of them are 
linearly dependent. For example, by the integral identity 
\eqref{eq:dWS_int}, we see that for $\lambda,\lambda'>0$ and 
$\lambda\neq\lambda'$, $\psi_{\lambda, k}(\tau)$and $\psi_{\lambda', k}(\tau)$ 
are only orthogonal if $\lambda-\lambda'\in2\bZ$. Moreover, the integral 
identity \eqref{eq:dWS_HJ} implies that for $\lambda>0$ and $\mu\geq0$
\begin{align}
\int_0^\infty \psi^\ast_{\lambda, k}(\tau) \psi_{\ii\mu, k}(\tau) 
\frac{\dd \tau}{\tau}
=\frac{-2\ii c^\ast_\lambda}{\pi(\mu^2+\lambda^2)} \Big( 
a_{\ii\mu} \ee^{\ii\pi\tfrac{\lambda-\ii\mu}{2}}
-b_{\ii\mu} \ee^{-\ii\pi\tfrac{\lambda-\ii\mu}{2}}
\Big).
\end{align}
This integral only vanishes if there is a relation between the allowed 
values of $\lambda$ and $\mu$. We encounter similar problems for the 
scalar product of $\psi_{\ii\mu, k}$ and $\psi_{\ii\mu', k}$. 

The mathematical reason for this is that the operator 
\eqref{eq:temporal_eigeneq1} is not essentially self-adjoint on 
$C_c^\infty(\bR_+)$ in the sense of $L^2\big(\bR_+,\frac{\dd \tau}{\tau}\big)$.
To find an orthonormal basis of (generalized) eigenfunctions, we need to choose 
a self-adjoint extension, which corresponds to specifying certain 
boundary conditions at $\tau=\infty$. Luckily, the theory of the 
operator \eqref{eq:temporal_eigeneq1} is well understood \cite{DeLee,Stempak}.\footnote{
Note that it is not what is commonly called the \emph{Bessel operator} 
$B$ but rather $\tau^2 B$. There is no simple operator-theoretic 
correspondence between self-adjoint realizations of $B$ and 
self-adjoint realizations of our temporal operator $\tau^2 B$. However, 
there is a formal \emph{transmutation formula} relating the resolvents 
of the two operators \cite{DeLee}.}
We discuss the details in Section \ref{sec:prop} and Appendix 
\ref{app:neg_exp}. In particular, 
the orthogonality of the mode functions corresponding to the 
self-adjoint realizations is proved in Appendix \ref{app:ortho}.

\subsection{Higher-spin valued eigenmodes}
\label{ssc:hs}
As indicated before, 
the operator algebra $\End(\cH)$ -- or at least its IR sector -- can be interpreted as quantized algebra of functions on an $S^2$ -bundle over spacetime. This internal sphere is parametrized by $T^i \sim t^i$, due to the constraint \eqref{t-y-constraints}. 
It is convenient to define the variables
\begin{align}
U^i := \frac{T^i}{Y^0}, 
\end{align}
which are normalized as $U_i U^i = \frac{1}{R^2r^2}$, and satisfy the commutation relations \eqref{TT-CR} and 
\begin{align}
[T^0,U^i] &= [T^0,\frac{T^i}{Y^0}] 
 = 0 \nn\\
%= \frac{1}{Y^0} (-\frac{i}{R}T^i) + [T^0,\frac{1}{Y^0}]T^i  \nn\\
% &= -\frac{i}{R} \frac{1}{Y^0} T^i + \frac{i}{R}\frac{1}{Y^0} T^i = 0  \nn\\
 [T^j,U^i] &= 0 = [U^j,U^i]  \ .
\end{align}
It follows that
\begin{align}
[T^\mu,Y^{lm}(U)] &= 0 =
\Box Y^{lm}(U)
\end{align}
for any polynomials $Y^{lm}(U)$ in the $U^i$.
In the semi-classical regime, this allows to organize
 the full space of modes in terms of higher-spin valued harmonics 
\begin{align}
 \Upsilon_{\Lambda,lm}(y^\mu,u)
= \psi_{\Lambda}(y) Y^{lm}(u),
\end{align}
where $Y^{lm}(u)$ are polynomials of order $l$ in  
\begin{align}
u^i = \frac{t^i}{y^0} = \frac{t^i}{\tau} , 
\qquad u_i u^i = \frac{1}{R^2r^2},
\end{align}
interpreted as spherical harmonics on the internal $S^2$. Since the $u^i$ transform as vector under $SO(3) \subset E(3)$, these modes should be interpreted as spin $l$ modes on spacetime, rather than as Kaluza-Klein modes.
We have thus obtained all higher-spin ($\hs$) eigenmodes of $\Box$, observing that
\begin{align}
 \Box \Upsilon_{\Lambda,lm}(y,u)
% &= \Box\psi_{lm}(y) Y^{lm}
% +\psi_{lm}(y)  \Box Y^{lm}  + 2[T^\mu,\psi_{lm}(y)] [T_\mu,Y^{lm}] \nn\\
&= \big(\Box\psi_{\Lambda}(y)\big) Y^{lm} \ .
\end{align}
This means that the $\hs$ modes do not acquire any 
intrinsic masses\footnote{This is nicer than for $k=-1$, where asymptotically small but non-vanishing masses are obtained \cite{SperlingStein,Manta:2024vol}.} .
We can finally define the quantization map \eqref{Q-def-general}  explicitly 
\begin{align}
\cQ(\Upsilon_{\Lambda,lm}) = \hat\Upsilon_{\Lambda,lm} \ 
\end{align}
as a unitary map compatible with $E(3)$, % (in analogy to Weyl quantization),
such that all quantum numbers are respected.

\section{Scalar field propagator}
\label{sec:prop}
 Any Green function $D_m(\tau,\tau',\vec{x},\vec{x}')$ 
of the Klein-Gordon equation, which respects the $E(3)$-invariance, 
may be written as a Fourier integral in the spatial variables.
Let $r := |\vec{x}-\vec{x}'|$, then formally
\begin{align}
D_m(\tau,\tau',\vec{x},\vec{x}') &=
\int  \ee^{\ii \vec{k}(\vec{x}-\vec{x}')}
\tilde{D}_m(\tau,\tau',|\vec{k}|) \frac{\dd^3\vec{k}}{(2\pi)^3}
\notag \\ \label{eq:Prop_as_Fourier}
&= \frac{1}{2\pi^2 r} \int_0^\infty  k \sin(kr)
\tilde{D}_m(\tau,\tau',k) \dd k,
\end{align}
where $\tilde{D}_m(\tau,\tau',k)$ is a Green function of
the one-dimensional operator
\begin{align} 
 P(\tau,\del_{\tau},k) + m^2, 
 \quad \label{eq:1d_op}
 P(\tau,\del_{\tau},k) :=  R^{-2} \Big( {\tau}^2 \del_{\tau}^2 
 + \tau  \del_{\tau} + {\tau}^2 k^2 \Big).
\end{align}
Not all of these Green functions arise from flat regime 
approximations of a propagator (i.e., have a reasonable 
interpretation as two-point function) associated to the 
matrix model. To have such an interpretation, we need 
to have an orthonormal basis of (generalized) eigenfunctions 
of \eqref{eq:1d_op} on $L^2(\bR_+, \tfrac{\dd \tau}{\tau})$. In 
mathematical terms, this means that the allowed Green functions 
should be the integral kernels of the boundary value of the 
resolvent of a \emph{self-adjoint} realization of the operator $P$ 
from \eqref{eq:1d_op}, 
\begin{align}
\tilde{D}_m(\tau,\tau',k) = (P(\cdot,\cdot,k) + m^2 -\ii0 )^{-1}(\tau,\tau').
\end{align}

Luckily, the self-adjoint realizations of \eqref{eq:1d_op} 
are well-understood \cite{Stempak,DeLee}. In terms of $\eta$ 
defined by $\tau = R\ee^{\eta}$, the one-dimensional operator 
reads
\begin{align}\label{eq:1d_SchrödOp} 
 P(\eta,\del_{\eta},k)
 =  R^{-2} \Big( \del_{\eta}^2 +  (kR)^2 \ee^{2\eta} \Big),
\end{align}
which is up to a global sign a Schrödinger operator with 
negative exponential. The 
self-adjoint realizations of $P(\eta,\del_\eta,k)$ are 
parametrized by a complex number $\gamma$ of unit absolute value, 
$|\gamma|=1$. They can thus be parametrized by $\alpha\in[0,2[$ 
if we write $\gamma=\ee^{\ii\pi\alpha}$. This parameter specifies 
the behavior of the eigenfunctions as $\eta\to\infty$, i.e., 
it corresponds to specifying boundary conditions that make 
\eqref{eq:1d_op} self-adjoint.

The spectrum of any of the self-adjoint realizations has a 
continuous and a discrete part. The (proper) eigenfunctions 
corresponding to the discrete 
spectrum are a distrecte subset of the $\psi_{\lambda,k}$ 
from \eqref{eq:eigenfunctions_lambda_imu}, while the 
(generalized) eigenfunctions corresponding to the continuous 
spectrum are a subset of the $\psi_{\ii\mu,k}$ from 
\eqref{eq:eigenfunctions_lambda_imu}.

\subsection{Timelike Green functions for self-adjoint boundary conditions}

The self-adjoint realizations $\{P^\alpha\;|\;\alpha\in[0,2[\}$ 
of the operator $P$ from \eqref{eq:1d_SchrödOp} are related to the 
self-adjoint realizations $M_{\ii l}^{\gamma=\ee^{\ii\pi\alpha}}(z,\del_z)$ 
of the Schrödinger operator with negative exponential potential discussed 
in Appendix \ref{app:neg_exp} via 
\begin{align}
P^\alpha(\eta,\del_\eta,k) 
:= - R^{-2} M^{\gamma=\ee^{\ii\pi\alpha}}_{\ii k R}(\eta,\del_\eta).
\end{align}

 The integral kernel of the resolvent 
 $T_k^{\alpha}(\nu^2;\tau,\tau')
 := (P^\alpha(\cdot,\cdot,k)-\nu^2)^{-1}(\tau,\tau') $ of 
 $P^\alpha(\tau,\del_\tau,k)$ can be expressed in terms of the 
 resolvent $R_{\ii l}^{\gamma=\ee^{\ii\pi\alpha}}(-\nu^2;z,z')$ 
 of $M_{\ii l}^{\gamma=\ee^{\ii\pi\alpha}}(z,\del_z)$, given 
 by \eqref{eq:resol_neg_exp}, via 
\begin{align}
T_k^{\alpha}(\nu^2;\tau,\tau') 
&= -\frac{R^2\pi\ii\; \theta({\tau'}-\tau)}{2(\ee^{\ii\pi \nu R}-\ee^{\ii\pi\alpha})}
  J_{R\nu}(k\tau) \big( \ee^{\ii\pi\nu R} H^{(+)}_{R\nu}( k \tau')
 + \ee^{\ii\pi\alpha} H^{(-)}_{R\nu}(k \tau') \big)
 \notag \\ \label{eq:cosmo_rsolvent}
 &\quad+(\tau \leftrightarrow \tau').
\end{align}
Correspondingly, the pertinent Green functions 
$\tilde{D}^\alpha_m(\tau,\tau',k)$ can be expressed as boundary 
values of $T_k^{\alpha}(\nu^2;\tau,\tau')$ via
\begin{align}
\tilde{D}^\alpha_m(\tau,\tau',k)
&= T_k^{\alpha}(-m^2+\ii0;\tau,\tau') 
\notag \\ 
&= \frac{- \pi\ii R^{2}}{2(\ee^{-\pi R m}-\ee^{\ii\pi\alpha})}
 \Big( \theta({\tau'}-\tau) J_{\ii R m}(k \tau) \big( \ee^{-\pi R m}
 H^{(+)}_{\ii R m}(k \tau')
 + \ee^{\ii\pi\alpha} H^{(-)}_{\ii R m}(k \tau') \big)
 \notag \\
 &\quad+\theta(\tau-{\tau'}) J_{\ii R m}(k \tau') \big( \ee^{-\pi R m }
 H^{(+)}_{\ii R m}(k \tau)
 + \ee^{\ii\pi\alpha} H^{(-)}_{\ii R m}(k \tau) \big)\Big).
 \label{eq:cosmo_Green_fct}
\end{align}

\subsection{The propagator as a mode integral}
From the resolvent of a self-adjoint operator, one can obtain 
the spectral projections via Stone's formula. This allows us to 
find the off-shell modes $\Upsilon^\alpha_{\lambda,\vec{k}}(\tau,\vec{x})$
corresponding to the self-adjoint realization  $P^\alpha(\tau,\del_\tau,k)$. 
It is clear that these self-adjoint realizations have a point spectrum 
because the prefactor 
\begin{align}
\frac{1}{\ee^{\ii\pi\nu R}-\ee^{\ii\pi\alpha}}
\end{align}
has poles if $\nu = \tfrac{\alpha+2 n}{R}$ for $n\in \bZ$. The 
corresponding eigenvalues are $\nu^2= \big(\tfrac{\alpha+2 n}{R}\big)^2$, 
$\alpha+2n > 0$, and the corresponding (proper and normalized) 
eigenfunctions are 
\begin{align}
\sqrt{2 (\alpha+2n)} J_{\alpha+2n}( k \tau), \quad \alpha+2n > 0.
\end{align}
Indeed, these eigenfunctions form an orthonormal subset of the 
$\psi_{\lambda,k}(\tau)$ from \eqref{eq:eigenfunctions_lambda_imu} because 
by the integral identity \eqref{eq:dWS_int}, we have 
\begin{align}
2\sqrt{(\alpha+2n)(\alpha+2m)} \int_0^\infty 
 J_{\alpha+2n}(\tau)  J_{\alpha+2m}(\tau) \frac{\dd \tau}{\tau}
 &= \delta_{m,n}.
\end{align}
Using Stone's formula, we also derive in Appendix \ref{app:mode_rep}
the (generalized) eigenfunctions corresponding to the continuous 
spectrum,
\begin{align}
\frac{\sqrt{\sinh(\pi\lambda )}}{2|\ee^{\ii\pi\alpha}-\ee^{-\pi\lambda }|}
 \Big(H^{(-)}_{\ii\lambda}(k \tau)
 + \ee^{-\pi(\lambda +\ii\alpha)}  H^{(+)}_{\ii\lambda}(k\tau)  \Big), 
  \quad \lambda > 0.
\end{align}
The orthogonality of all eigenfunctions is proved in Appendix 
\ref{app:ortho}. By the spectral theorem (cf. also the representation 
\eqref{eq:resolvent_sa_mode_sum}), the Green function 
\eqref{eq:cosmo_Green_fct} can be represented as a mode sum/integral, 
\begin{align}
&\quad \tilde{D}^\alpha_m(\tau,\tau',k) \notag \\
&= R^2 \int_0^\infty  
\frac{\Big(H^{(-)}_{\ii\lambda }(k \tau)
 + \ee^{-\pi(\lambda +\ii\alpha)}  H^{(+)}_{\ii\lambda }(k\tau)  \Big)
 \Big(H^{(-)}_{\ii\lambda }(k \tau')
 + \ee^{-\pi(\lambda +\ii\alpha)}  H^{(+)}_{\ii\lambda }(k\tau')  \Big)^\ast
 }{-\lambda^2+(mR)^2-\ii0} 
 \notag \\ &\hspace{12ex}\times 
  \frac{\lambda \sinh(\pi\lambda )}{2|\ee^{\ii\pi\alpha}-\ee^{-\pi\lambda }|^2} 
  \dd \lambda
 \notag \\ \label{eq:1d_GF_modes}
 &\quad + R^2 \sum_{n=0}^\infty \frac{2(\alpha+2n) 
 J_{\alpha+2n}(k\tau)J_{\alpha+2n}^\ast(k\tau')}{(\alpha+2n)^2+ (mR)^2}.
\end{align}
Note that $J_{\nu}(z) = J_{\nu}^\ast(z)$ for $\nu, z >0$. 
Summing up, we find the following off-shell modes corresponding 
to the continuous and discrete spectrum, respectively:
\begin{empheq}[box=\widefbox]{align}
\Upsilon_{\vec{k},\lambda}^{(\alpha)}(\tau,\vec{x}) 
&= \frac{R\sqrt{\lambda\sinh(\pi\lambda)}}{4\pi^{\frac{3}{2}}
|\ee^{\ii\pi\alpha}-\ee^{-\pi\lambda }|}
\ee^{\ii \vec{k}\vec{x}} \big(H^{(-)}_{\ii\lambda }(k \tau)
 + \ee^{-\pi(\lambda +\ii\alpha)}  H^{(+)}_{\ii\lambda }(k\tau)  \big); 
 \notag \\
 \tilde{\Upsilon}_{\vec{k},n}^{(\alpha)}(\tau,\vec{x})
 &= \frac{R\sqrt{\alpha+2n}}{2\pi^{\frac{3}{2}}} \ee^{\ii \vec{k}\vec{x}}
  J_{\alpha+2n}(k\tau), \notag \\
&\word{where} \lambda  > 0 \word{and}  n\in\bN_0,\;\alpha+2n> 0.
\end{empheq}
They satisfy 
\begin{align}
\Box\Upsilon_{\vec{k},\lambda}^{(\alpha)} 
&= -\big(\tfrac{\lambda}{R}\big)^2 \Upsilon_{\vec{k},\lambda}^{(\alpha)} , 
\notag \\ 
\Box  \tilde{\Upsilon}_{\vec{k},n}^{(\alpha)}
&= \big(\tfrac{\alpha+2n}{R}\big)^2 \tilde{\Upsilon}_{\vec{k},n}^{(\alpha)}.
\end{align}

The propagator therefore reads 
\begin{empheq}[box=\widefbox]{align}
D_m^{\alpha}(\tau,\tau',\vec{x},\vec{x}') 
&= \int_{\bR^3} \Bigg(\int_0^\infty  
\frac{\Upsilon_{\vec{k},\lambda}^{(\alpha)}(\tau,\vec{x}) 
\Upsilon_{\vec{k},\lambda}^{(\alpha) \ast}(\tau',\vec{x}')
}{-\lambda^2+(mR)^2-\ii0} \dd \lambda 
\notag \\ \label{eq:propagator_as_mode_int}
&\hspace{9ex} 
+\sum_{n=0}^\infty \frac{ \tilde{\Upsilon}_{\vec{k},n}^{(\alpha)}(\tau,\vec{x})
 \tilde{\Upsilon}_{\vec{k},n}^{(\alpha)\ast}(\tau',\vec{x}')}{(\alpha+2n)^2+ (mR)^2}
\Bigg)\dd^3\vec{k}.
\end{empheq}

\subsection{Exact evaluation of the Fourier integral}
We can determine the integral \eqref{eq:Prop_as_Fourier} exactly 
if we insert $\tilde{D}^\alpha_m(\tau,\tau',k)$ for 
$\tilde{D}_m(\tau,\tau',k)$. Using the connection formulas 
\eqref{eq:connect_KH} and \eqref{eq:connect_IJ} to express 
the Bessel and Hankel functions in terms of modified Bessel 
functions of the first and second kind, we obtain
\begin{align}
  &\quad \tilde{D}^\alpha_m(\tau,\tau',k)\notag \\
&=   \frac{ R^{2}\theta({\tau'}-\tau) }{(\ee^{- \pi R m}-\ee^{\ii\pi\alpha})}
  \big( \ee^{\ii\pi\alpha}I_{\ii R m}(\ii k \tau) K_{\ii R m}(\ii k \tau') -\ee^{-\pi R m}
 I_{\ii R m}(-\ii k \tau) K_{\ii R m}(-\ii k \tau')\big)
 \notag \\
 &\quad+(\tau\leftrightarrow \tau').
\end{align}
The propagator is then the distributional boundary value 
of the Fourier integral \eqref{eq:Prop_as_Fourier}. 
This integral can be determined using the identity \eqref{eq:KIsin_int}, 
which implies 
\begin{align}
 &\quad\lim_{\eps\downarrow0}\int_0^\infty   k \sin(kr)\;\big(
 \ee^{\ii\pi\alpha} I_{\ii R m}(\ii k \tau) K_{\ii R m}(\ii k (\tau'-\ii\eps)) \notag \\
 &\hspace{9ex}-\ee^{-\pi R m}
 I_{\ii R m}(-\ii k \tau) K_{\ii R m}(-\ii k (\tau'+\ii\eps))\big)\dd k \notag \\
 &= -\ii \frac{\sqrt{\pi}\Gamma\big(\tfrac{3}{2}+\ii Rm\big)}{2^{\ii Rm}}
  \frac{r}{(2\tau\tau')^{\tfrac{3}{2}}} \Big( \ee^{\ii\pi\alpha} {\bf Z}_{1,\ii Rm}(w-\ii0)
  +\ee^{-\pi R m}  {\bf Z}_{1,\ii Rm}(w+\ii0) \Big),
  \label{eq:important_integral}
\end{align}
where ${\bf Z}_{\mu,\nu}(z)$ is the Gegenbauer function \eqref{eq:def_Z} and
where
\begin{align}
 w := \frac{\tau^2+{\tau'}^2-r^2}{2\tau\tau'} = 1+ \frac{(\tau-{\tau'})^2-r^2}{2\tau\tau'}
 = -1+ \frac{(\tau+{\tau'})^2-r^2}{2\tau\tau'}.
\end{align}
By the symmetry of \eqref{eq:important_integral} under the exchange of
$\tau$ and $\tau'$, we obtain an explicit formula for the propagator 
corresponding to boundary conditions specified by $\alpha$:
% \begin{align}\label{eq:Feynman_all_candidates}
%  D_{m}^{\alpha}
%  = \frac{ -\ii R^{2}\Gamma\big(\tfrac{3}{2}+\ii Rm\big) 
%  }{2^{1+\ii Rm}(\ee^{- \pi R m}-\ee^{\ii\pi\alpha})}
%   \frac{\ee^{\ii\pi\alpha} {\bf Z}_{1,\ii Rm}(w-\ii0)
%   +\ee^{-\pi R m}  {\bf Z}_{1,\ii Rm}(w+\ii0)}{(2\pi \tau\tau')^{\tfrac{3}{2}}}.
% \end{align}
% We may rewrite this as 
\begin{align}
D_{m}^{\alpha}
 = \frac{ \ii R^{2}\Gamma\big(\tfrac{3}{2}+\ii Rm\big) 
 }{2^{1+\ii Rm}(2\pi \tau\tau')^{\tfrac{3}{2}}}
 &\Bigg({\bf Z}_{1,\ii Rm}(w-\ii0)
 \notag \\
&  -\frac{\ee^{-\pi R m}}{\ee^{- \pi R m}-\ee^{\ii\pi\alpha}} 
  \Big( {\bf Z}_{1,\ii Rm}(w+\ii0)
  +{\bf Z}_{1,\ii Rm}(w-\ii0)\Big)
  \Bigg).\label{eq:Feynman_all_candidates}
\end{align}
The first term in \eqref{eq:Feynman_all_candidates} does 
not depend on the boundary conditions and yields the Dirac 
delta upon acting with $\Box+m^2$ on $D^\alpha_m$. 
The second term, which contains all information about the 
boundary conditions, is a solution of the Klein-Gordon equation. 
The prefactor $\tfrac{\ee^{-\pi R m}
}{\ee^{- \pi R m}-\ee^{\ii\pi\alpha}}$ indicates that the 
dependence on the choice of boundary conditions is suppressed 
in the semi-classical regime, where $\tau$, $\tau'$ and $m$ 
become large. We validate the latter observation more carefully 
in the following subsection.

\subsection{Asymptotic behavior in the semi-classical regime}
\label{ssc:semi_classical}
Although we are able to derive exact formulas for the propagators 
$D^\alpha_m(\tau,\tau',\vec{x},\vec{x}')$, they are expeceted to be a good 
approximation to the prediction of the matrix model only in the 
semi-classical regime, i.e., in a local patch at late times. 
Notice that the constant physical mass $m_0$ is related to the 
mass $m$ via the dilaton, $m^2= \rho^2 m_0^2$. Since the dilaton 
$\rho$ increases with time, a local patch in the late time regime 
is subject to the assumptions 
\begin{align} \label{eq:conditions_late_time}
\tau,\tau',m\gg1,\quad |\tau - \tau'| \ll \tau,\tau',m.
\end{align}

We determine the asymptotic behavior in this regime as 
follows. First, we derive the asymptotic behavior of 
the timelike Green functions. Then, we argue that the 
contribution that depends on the boundary conditions is 
suppressed in the Fourier integral \eqref{eq:Prop_as_Fourier} 
due to rapid oscillatory behavior. 

This procedure seems optimal because 
\begin{itemize}
\item[(1)] the asymptotic behavior of the exact solution 
\eqref{eq:Feynman_all_candidates} is difficult to determine 
because it involves a simultaneous approximation of the 
Gegenbauer function ${\bf Z}_{1,\ii Rm}(w\pm\ii0)$ near its 
poles at $w=\pm1$ and for large imaginary order $\ii Rm$; 
\item[(2)] the flat regime limit of the modes in 
\eqref{eq:propagator_as_mode_int} causes both the sum over $n$ 
and the integral over $\lambda$ to diverge due to the asymptotics 
\eqref{eq:asymp_J} and \eqref{eq:asymp_H} of the Bessel functions. 
    Thus, one should perform the sum over $n$ and the integral 
    over $\lambda$ before investigating the asymptotic behavior. 
    This naturally leads to the described procedure. 
\end{itemize}

We start by writing the timelike Green function 
\eqref{eq:cosmo_Green_fct} as 
\begin{align}\label{eq:cosmo_GF_nu_asymp}
\tilde{D}^\alpha_m
&= \frac{- \pi\ii R^{2}}{2(\ee^{-\pi \nu }-\ee^{\ii\pi\alpha})}
 \Big( \theta({\tau'}-\tau) J_{\ii \nu }(\nu z) \big( \ee^{-\pi \nu}
 H^{(+)}_{\ii \nu}(\nu z')
 + \ee^{\ii\pi\alpha} H^{(-)}_{\ii \nu}(\nu z') \big)
 \notag \\
 &\quad+\theta(\tau-{\tau'}) J_{\ii \nu}(\nu z') \big( \ee^{-\pi \nu }
 H^{(+)}_{\ii \nu}(\nu z)
 + \ee^{\ii\pi\alpha} H^{(-)}_{\ii \nu}(\nu z) \big)\Big),
\end{align}
where $\nu = R m$, $z=\frac{k \tau}{R m}$ and  $z'=\frac{k \tau'}{R m}$. 
Then we assume that $\nu$ is large, while keeping $z$ and $z'$, 
which are ratios of time and mass, arbitrary. These asymptotics 
are known \cite{Dunster90}. For convenience of the reader, we 
present them in Appendix \ref{app:Bessel}. 

Before we derive the desired asymptotics of \eqref{eq:cosmo_GF_nu_asymp}, 
we note that the timelike operator can be written as 
\begin{align}
\tau^2 \Big( \del_\tau^2 + \frac{1}{\tau}\del_\tau + k^2 + \frac{m^2}{\tau^2}\Big).
\end{align}
One should therefore expect that the on-shell energy is roughly 
$E^2=E^2(\tau)=k^2 + \frac{m^2}{\tau^2}$ in this late time regime.
The quantity $E$ should thus appear in the asymptotic form of the 
Green function. 

If we insert the asymptotic behavior \eqref{eq:H_asymp_inu} and 
\eqref{eq:J_asymp_inu} into \eqref{eq:cosmo_GF_nu_asymp}, we obtain 
\begin{align}
\tilde{D}^\alpha_m 
&\sim 
 \frac{-  R^{2} \theta({\tau'}-\tau) 
 }{2\nu (1+z^2)^{\frac{1}{4}}(1+{z'}^2)^{\frac{1}{4}}(\ee^{-\pi \nu }-\ee^{\ii\pi\alpha})} 
\notag \\ &\quad\times
 \Big( 
 \ee^{\ii \nu (\xi(z')+\xi(z))} + \ii \ee^{\ii\pi\alpha} 
 \ee^{-\ii \nu(\xi(z')-\xi(z))}
 +\ee^{-\pi\nu} \big( \ii \ee^{\ii \nu(\xi(z')-\xi(z))} 
 -\ee^{\ii\pi\alpha} \ee^{-\ii \nu (\xi(z')+\xi(z))} \big) \Big)
\notag \\ 
 &\quad+(\tau\leftrightarrow \tau')
\end{align}
for $\nu\gg1$. The terms with a prefactor $\ee^{-\pi\nu}$ in the 
last equation are suppressed. Moreover, the terms that only depend 
on $\xi(z)+\xi(z')$ remain unchanged under the exchange 
$(\tau\leftrightarrow \tau')$. We may thus further simplify 
\begin{align}
\tilde{D}^\alpha_m 
&\sim\frac{ R^{2} }{2\nu (1+z^2)^{\frac{1}{4}}(1+{z'}^2)^{\frac{1}{4}}} 
\notag \\ &\quad\times
 \Bigg( \ee^{-\ii\pi\alpha} \ee^{\ii \nu (\xi(z')+\xi(z))} 
 + \ii \Big( \theta({\tau'}-\tau)  \ee^{-\ii \nu(\xi(z')-\xi(z))}
 +  (\tau\leftrightarrow\tau')\Big) 
 \Bigg).
\end{align}
If we write $E':=E(\tau')$, we have 
$\nu (1+z^2)^{\frac{1}{4}}(1+{z'}^2)^{\frac{1}{4}} 
= \sqrt{E\tau} \sqrt{E'\tau'}$ and 
\begin{align}
\nu \xi(z) = E\tau + mR \ln\Big( \frac{k\tau}{mR+ E \tau}\Big).
\end{align}
Thus, 
\begin{align}\label{eq:approx_GF_yym}
\tilde{D}^\alpha_m 
&\sim\frac{ R^{2} }{\sqrt{\tau\tau'}}
\Bigg( \frac{\ee^{-\ii\pi\alpha} \ee^{\ii ( E\tau+E'\tau')}
 \big( \frac{k^2 \tau\tau'}{(mR+E\tau)(mR+E'\tau')}\big)^{\ii mR}
}{2\sqrt{EE'}}
\notag \\
&\qquad+ \ii
\frac{\theta({\tau'}-\tau)  \ee^{-\ii ( E'\tau'-E\tau)}
 \big( \frac{\tau (mR+E\tau')}{\tau' (mR+E\tau)}\big)^{\ii mR} 
 + (\tau\leftrightarrow \tau')}{2\sqrt{EE'}} \Bigg).
\end{align}

The approximation \eqref{eq:approx_GF_yym} is valid if 
$\tau,\tau',m\gg1$ and follows from the asymptotic behavior 
\eqref{eq:H_asymp_inu} and \eqref{eq:J_asymp_inu} of the 
Bessel functions. Using the second condition for the late 
time regime, $|\tau - \tau'| \ll \tau,\tau',m$, it may be further simplified. 
Under the latter assumption, we have $E\sim E'$ and 
 $\frac{\tau (mR+E\tau')}{\tau' (mR+E\tau)}\sim 1$. We have numericaly verified 
using Wolfram Mathematica that 
\begin{align}
\tilde{D}^\alpha_m 
&\sim\frac{ R^{2} }{\sqrt{\tau\tau'}}
\Bigg( \frac{\ee^{-\ii\pi\alpha} \ee^{\ii  \tilde{E}(\tau+\tau')}
 \big( \frac{k^2 \tau\tau'}{(mR+E\tau)(mR+E'\tau')}\big)^{\ii mR}
}{2\tilde{E}}+ \ii
\frac{\theta({\tau'}-\tau)  \ee^{-\ii \tilde{E} (\tau'-\tau)}
 + (\tau\leftrightarrow \tau')}{2\tilde{E}} \Bigg)
 \label{eq:approx_GF_FR}
\end{align}
if $\tau,\tau',m\gg1$ and $|\tau - \tau'| \ll \tau,\tau',m$. 
Here, $\tilde{E}:= \sqrt{k^2+\frac{(mR)^2}{\tau\tau'}}$.
We then use the distributional identity
\begin{align}
\frac{\ee^{-\ii q t}}{2q} 
= -\ii  \int_{-\infty}^\infty \frac{\dd p}{2\pi} 
\frac{\ee^{-\ii p t}}{ -p^2+q^2-\ii0},\quad q,t>0,
\end{align}
to write the approximation of the propagator in the semi-classical 
regime as 
 \begin{empheq}[box=\widefbox]{align}
&\quad D_m^\alpha(\tau,\tau',\vec{x},\vec{x}') 
\notag \\
&\sim  \frac{ R^{2}}{ \sqrt{\tau\tau'}}
 \Bigg( 
 \int 
    \frac{ \ee^{-\ii k^0 (\tau-\tau') +\ii \vec{k} (\vec{x}-\vec{x}')}}{k^\mu k_\mu +\frac{m^2}{\tau\tau'} -\ii0}
 \frac{\dd^4k}{ (2\pi)^4}
 \notag \\ &\quad +
 \ee^{-\ii\pi\alpha}
 \int \frac{\ee^{\ii \tilde{E} (\tau+\tau') +\ii \vec{k} (\vec{x}-\vec{x}')}}{2\tilde{E}}
 \Big( \tfrac{k^2 \tau\tau'}{(mR+E\tau)(mR+E'\tau')}\Big)^{\ii mR}
 \frac{\dd^3\vec{k}}{ (2\pi)^3} \label{eq:FR_propa}
\Bigg).
\end{empheq}
Up to a time-dependent prefactor, the first line of \eqref{eq:FR_propa}, 
which is universal in the sense that it does not depend on the choice of 
boundary conditions, looks like a Feynman propagator of a scalar field 
in Minkowski space with a mass $\frac{m^2}{\tau\tau'}$. 
The second line of \eqref{eq:FR_propa}, which encodes all information 
about the boundary conditions in the simple prefactor $\ee^{-\ii\pi\alpha}$, 
looks like a modification of a "reflected" Wightman function, where 
we mean by "reflected" that the exponential depends on $\tau+\tau'$, and where 
the modification is the factor 
\begin{align}
 \Bigg( \frac{k^2 \tau\tau'}{(mR+E\tau)(mR+E'\tau')}\Bigg)^{\ii mR},
\end{align}
again up to a time-dependent prefactor.

The reflected part is rapidly oscillating in the semi-classical 
regime, where $\tau+\tau'\gg1$. Hence, it is suppressed compared to the 
universal part. Then the dependence on the choice of boundary 
conditions is also suppressed, and we recover the standard form  of the propagator
\begin{empheq}[box=\widefbox]{align}
D_m^\alpha(\tau,\tau',\vec{x},\vec{x}') 
&\sim D_m^{\rm semi-cl}(\tau,\tau',\vec{x},\vec{x}') 
:= \frac{ R^{2}}{ \sqrt{\tau\tau'}}
 \int 
    \frac{ \ee^{-\ii k^0 (\tau-\tau') +\ii \vec{k} (\vec{x}-\vec{x}')}}{k^\mu k_\mu +\frac{m^2}{\tau\tau'} -\ii0}
 \frac{\dd^4k}{ (2\pi)^4} \ .
 \label{eq:semiclassical_prop}
\end{empheq}
We remark that we have numerically validated the suppression 
of the reflected part. The numerical evaluation needs to be 
performed with high precission in order to accurately account 
for the rapid oscillations.

\section{Discussion}
In this work, we have described an expanding, spatially flat ($k=0$) 
FLRW quantum spacetime with a Big Bang, which may emerge in the semi-classical 
limit from the IKKT matrix model. Similar to previous studies of open $k=-1$ 
FLRW quantum spacetimes with a Big Bounce \cite{BStein}, our results should be 
viewed in the context of emergent spacetime and gravity within the IKKT framework, 
which can be viewed as an alternative approach to string theory \cite{IKKT97}. 
The algebraic description of the new quantum spacetime is remarkably simple, 
so that it should provide a useful starting point for further work on the emergent 
physics on quantum spacetime.

As a first step in this direction,
we have explicitly described the quantization of scalar fields 
and higher-spin modes. Notably, the quantization of higher-spin modes 
in the $k = 0$ case turned out to be simpler than in the $k = -1$ case. 
Moreover, the higher-spin modes do not acquire 
any intrinsic masses in the $k = 0$ case, and can be fully described in terms of scalar modes 
and standard spherical harmonics on the internal $S^2$.

We also found that the semi-classical d'Alembertian is related to 
the d'Alembertian on the Poincaré patch of de Sitter space by 
a similarity transformation. Thus, it is exactly solvable
--- cf. for example \cite{BirrellDavies,FSS13}. However, this 
operator is not essentially self-adjoint and there is a one-parameter 
family of self-adjoint realizations, which we traced back to the 
self-adjoint realizations of a Schrödinger operator with negative 
exponential potential \cite{DeLee,Stempak}. All these self-adjoint 
realizations stand on an equal footing, and any of them allows us 
to express the scalar field propagator as a mode integral. The 
different realizations correspond to different boundary conditions 
at temporal infinity. That is, they arise from global properties of 
the semi-classical geometry and d'Alembertian. 

However, the semi-classical approximation is only applicable 
in a small patch at late times, where global properties of the 
semi-classical geometry should play no important role. Indeed, we 
showed that the dependence on the choice of boundary conditions is 
suppressed in the regime where the approximation is valid.

We observed interesting analogies, but also interesting differences
between the current $k = 0$ FLRW case and the $k = -1$ FLRW case 
described in \cite{BStein}. In both cases, 
the effective scalar field propagator can be split into two parts: 
one that depends on the \emph{difference} of the time variables 
and one that depends on the \emph{sum} of the time variables.  
This splitting reflects the presence of a cosmic singularity.
The part that depends on the difference of the time variables 
resembles the propagator of a scalar field in Minkowski space 
in both cases. 
The part that depends on the sum of time variables 
exhibits different properties in the two cases. In the $k=-1$ case, 
it describes pre-post-Big-Bounce correlations, which become large 
if the sum of the time variables is small. In the current 
case where $k=0$, it can be interpreted in terms of a reflection at the Big Bang,
which is suppressed in the semi-classical regime.
Its explicit form encodes the dependence of the semi-classical 
propagator on the boundary conditions 
at time-like infinity.

\appendix
\section{Algebraic identities}
\label{app:group-id}
The generators $M^{ab}, \ a,b=0,...,5$ of the doubleton 
representations $\cH_n$ of $\mso(4,2)$ satisfy the following 
identity \cite{SperlingStein, Steinbook}:
\begin{align}
     \sum_{a,b= 0,1,2,3,4,5}  \eta_{ab} M^{ac} M^{bd} + (c\leftrightarrow d) =
\frac{1}{2}(n^2-4) \eta^{cd} = 2 \frac{R^2}{r^2} \eta^{cd}  \,.
 \label{MM-contraction-42}
\end{align}
This implies the constraints \eqref{T-Y-constraints} noting that 
$\beta_a \beta^a = 0 = \alpha_a \beta^a$.

Let us derive the commutation relations \eqref{Yi-Yj-CR}, 
valid in the case $n=0$. To this end, we want to express 
$M^{ij}$ in terms of $Y^\mu$, $T^\mu$ and $X^4$. More generally, 
we can describe all $M^{\mu\nu}$ solely as a function of $T^\mu$, 
$Y^\mu$ and $X^4$, starting from the explicit representation 
\cite{MantaStein25}
\begin{align}\label{eq:Mmunu1}
M^{\mu\nu} = r X_4^{-1} \Big( M^{\mu 4} M^{\nu 5} - M^{\nu 4} M^{\mu 5}\Big),
\end{align}
which translates to
\begin{align}
M^{\mu\nu} &= X_4^{-1} \Big(\big( RT^\mu - M^{\mu 0} \big) 
\big(Y^\nu-\delta^{\nu0} X^4\big)
-\big( RT^\nu - M^{\nu 0} \big) 
\big(Y^\mu-\delta^{\mu0} X^4\big)\Big).
\end{align}
By \eqref{eq:Mmunu1}, we have 
\begin{align}
X_4 M^{\mu 0} + M^{\mu 0}  (Y^0-X^4)  
=   R \Big(T^\mu (Y^0-X^4) 
-  T^0 (Y^{\mu}-\delta^{\mu 0} X^4) \Big)
\end{align}
We can solve this for $M^{\mu 0}$ (or rather $M^{i 0}$) by using 
$[M^{\mu\nu},X^4]=0$. This yields 
\begin{align}\boxed{ 
 M^{i 0} 
 = R \Big(T^i (Y^0-X^4) -  T^0 Y^{i} \Big) {(Y^0)}^{-1}.}
\end{align}
Consequently,
\begin{align}
M^{ij} 
&= RX_4^{-1} \Big( \big( T^i X^4 
+  T^0 Y^{i}\big) {(Y^0)}^{-1}  
Y^j - (i\leftrightarrow j) \Big).   
% \nn\\
% &\harold{  =R\big(1 + i r^2 R X_4^{-1} (Y^0)^{-1} T^0\big)(T^i Y^j - T^j Y^i)(Y^0)^{-1}
% }
\end{align}

Next, note that the operator $T^i Y^j - T^j Y^i$ commutes both 
with $X^4$ and $Y^0$. Then, using the notation  $O \coloneqq R  X_4^{-1} 
\big( -\tfrac{\ii}{R}\one + T^0\big) $ as well as the commutation relations 
\eqref{Y0-Yi-CR} and \eqref{X4-T-CR}, we find 
\begin{align}
M^{ij} &= R(Y^0)^{-1} (T^i Y^j - T^j Y^i) 
+ O  \big(Y^i(Y^0)^{-1} Y^j - Y^j(Y^0)^{-1} Y^i  \big)
\notag \\
&=  R(Y^0)^{-1} (T^i Y^j - T^j Y^i) 
+ O (Y^0)^{-1}  \Big([Y^i, Y^j] + \big( [Y^0,Y^i] Y^j-[Y^0,Y^j] Y^i \big)(Y^0)^{-1} \Big)
\notag \\
&=  R \big(1+\ii r^2 O (Y^0)^{-1}  \big) (T^i Y^j - T^j Y^i) (Y^0)^{-1}
- \ii r^2 O (Y^0)^{-1}  M^{ij}.
\end{align}
Then the factor $1+\ii r^2 O (Y^0)^{-1}$ drops out and we find 
\begin{align}\label{eq:Mij_TY}
\boxed{
M^{ij} = R(Y^0)^{-1} (T^i Y^j - T^j Y^i) .
}
\end{align}

\section{Invariant volume form}
\label{sec:invar-vol-form}

Given some quantized symplectic space with quantization map $\cQ$, the trace 
over (suitable) operators in $\End(\cH)$ can typically be related to the 
integral over $\cM$ w.r.t. the symplectic volume form \cite{Steinbook}, 
\begin{align}
\Tr\big(\cQ(f)\big) \sim \int_{\cM} \Omega f \word{in the semiclassical regime.}
\end{align}
If we insert $f= \phi^\ast (\square+m^2) \phi$, where $\square$ is the differential 
operator $-\{t^\mu,\{t_\mu,\cdot\}\}$ from \eqref{eff-dAlembertian} and where 
$\phi$ is a scalar field, we obtain the quantization \eqref{eq:matrix_action} of the 
action given by \eqref{eq:semicl_action}. For scalar fields, the integral over the internal 
$S^2$ is trivial and thus, we effectively have to deal with an integral over 
the four-dimensional spacetime $\cM^{1,3}$ with respect to an induced volume form 
$\Omega^{1,3}$. 

This volume form on the spacetime $\cM^{1,3}$ can be described as pull-back 
of the unique $SO(4,1)$-invariant volume form on the 4-hyperboloid $H^4\subset \bR^{4,1}$ 
spanned by the $x^a$ generators due to \eqref{XX-R-fuzzy-H4}. It can be 
written explicitly as \cite{SperlingStein}
\begin{align}
\Omega^{1,3} 
&= \frac{R^3}{x^4} \dd x^0 \dots \dd x^3.
\end{align}
Equivalently, we may write
\begin{align}
\Omega^{1,3} = \frac{R^3}{x^0} \dd x^4 \dd x^1 \dots \dd x^3
 = \frac{R^3}{y^0} \dd y^0 \dd x^1 \dots \dd x^3 
\label{eq:symplectic_form}
 =: \rho_M(y^0)  \dd y^0 \dd x^1 \dots \dd x^3,
\end{align}
where $y^0 := x^0 + x^4$ and where 
\begin{align}
 \rho_M(y^0) := \frac{R^3}{y^0}
\end{align}
is the symplectic density. The equivalence of the first form of 
\eqref{eq:symplectic_form} to the original form follows from
\begin{align}
\label{xdx-rel}
x^4 \dd x^4 \dd x^1 \dd x^2 \dd x^3 = x^0 \dd x^0\dd x^1 \dd x^2 \dd x^3,
\end{align}
which in turn follows from the $SO(4,1)$-invariant constraint 
$-{x^0}^2+ \sum_{i=1}^4 {x^i}^2=0$.
The second form of \eqref{eq:symplectic_form} turns out to be the 
most natural form to describe a $k=0$ FLRW spacetime. 
Its equivalence to the other two forms again follows from \eqref{xdx-rel}, 
which implies
\begin{align} \label{eq:sympl_form_equiv_aux}
(x^0 + x^4) \dd x^4 \dd x^1 \dots \dd x^3
 &= x^0 ( \dd x^0 +\dd x^4) \dd x^1 \dots \dd x^3.
\end{align}
To obtain the third equality in \eqref{eq:symplectic_form}, we have to 
divide both sides of \eqref{eq:sympl_form_equiv_aux} by $x^0(x^0+x^4)$.

\section{Special functions}
\label{app:Bessel}

In this appendix we list the pertinent properties of the special 
functions that appear in our derivation of the scalar field 
propagator.

\subsection{Bessel functions and modified Bessel functions}
The \emph{Bessel equation} is
\begin{align}
\label{eq:Bessel}
\Big(\frac{\dd^2}{\dd z^2} + \frac{1}{z} \frac{\dd }{\dd z}
+ 1 - \frac{\nu^2}{z^2}\Big) f(z)=0.
\end{align}
If $\nu\notin\{0,-1,-2,\dots\}$, then $J_{\pm\nu}(z)$ is a pair of
linearly independent solutions with definite behavior
$\sim\big(\frac{z}{2}\big)^{\pm\nu}$ at $z=0$ \cite{NIST}, where
\begin{align} \label{eq:defJ}
J_\nu(z) =
\sum_{k=0}^\infty (-1)^k \frac{\big(\tfrac{z}{2}\big)^{2k+\nu}}{k!\Gamma(1+\nu+k)}.
\end{align}
 One may introduce the \emph{Bessel function of the second kind}
 \begin{align}
 Y_\nu(z) := \frac{J_\nu(z) \cos\pi\nu - J_{-\nu}(z)}{\sin\pi\nu},
 \end{align}
defined for $\nu\in\{0,-1,-2,\dots\}$ by application of the rule of
de l'Hôpital. $J_\nu(z)$ and $Y_\nu(z)$ are linearly independent
for any value of $\nu$.

A pair of linearly independent solutions with
definite behavior as $z\to\infty$ are the \emph{Hankel functions}
\begin{align}
\label{eq:def_Hankel}
H^{(1/2)}_\nu(z) := J_\nu(z) \pm\ii Y_\nu(z) =: H^{(\pm)}_\nu(z),
\end{align}
with
\begin{align}
\label{eq:asymp_H}
H_\nu^{(\pm)}(z)\sim \sqrt{\frac{2}{\pi z}}
\ee^{\pm\ii\big( z-\frac{\pi\nu}{2} - \frac{\pi}{4} \big)}
\word{as} z\to\infty \text{ if } -\pi+\eps \leq \arg(z) \leq 2\pi-\eps.
\end{align}
The following connection formulas are important for our purpose \cite{NIST}:
\begin{align}
 \label{eq:connect_H_pmnu}
 H_{-\nu}^{(\pm)}(z) &= \ee^{\pm\pi\ii \nu} H_{\nu}^{(\pm)}(z),
 \\ \label{eq:connect_JH}
 J_\nu(z) &= \frac12 \Big( H_\nu^{(+)} + H_\nu^{(-)} \Big),
 \\ \label{eq:connect_HJ}
  H_{\nu}^{(\pm)}(z) &= \frac{\pm\ii}{\sin(\pi\nu)} 
  \Big( \ee^{\mp\ii\pi\nu} J_\nu(z)
  - J_{-\nu}(z)\Big).
\end{align}
For fixed $\nu$, the Bessel function $J_\nu(z)$ has the 
asymptotics \cite{NIST}
\begin{align}
J_\nu(z) &\sim \sqrt{\frac{2}{\pi z}} 
\cos\big(z-\tfrac{\pi\nu}{2}-\tfrac{\pi}{4}\big) 
\word{as} z\to\infty \textup{ if } |\arg(z)|\leq \pi-\delta.
\label{eq:asymp_J}
\end{align}

The Hankel functions of imaginary order have the following 
asymptotic behavior for $z>0$ fixed \cite{Dunster90}: 
\begin{align}
H^{(\pm)}_{\ii\nu}(\nu z) 
&\sim \Big( \frac{2}{\pi \nu \sqrt{1+z^2}} \Big)^{\frac{1}{2}}
\ee^{\pm\ii \big( \nu \xi(z) - \tfrac{\pi\ii\nu}{2} - \tfrac{\pi}{4} \big)}
\word{as} \nu\to\infty.
\label{eq:H_asymp_inu}
\end{align}
Here, $\xi(z)$ is the non-trivial function 
\begin{align}
\xi(z)= \sqrt{1+z^2} + \ln\Big( \frac{z}{1+\sqrt{1+z^2}}\Big).
\end{align}
Notice that $\xi(z)$ is monotonically increasing with $\xi(z)\to-\infty$
as $z\to 0$ (the logarithm is dominating for small $z$) and 
$\xi(z)\to+\infty$ for $z\to\infty$ (the square root is dominating 
for large $z$).
The connection formula \eqref{eq:connect_HJ} then yields 
\begin{align}
J_{\ii\nu}(\nu z)
&\sim \Big( \frac{2}{\pi \nu \sqrt{1+z^2}} \Big)^{\frac{1}{2}}
\cos\big( \nu \xi(z) - \tfrac{\pi\ii\nu}{2} - \tfrac{\pi}{4} \big)
\word{as} \nu\to\infty.
\label{eq:J_asymp_inu}
\end{align}

The \emph{modified Bessel equation}, obtained by replacing
$z$ with $\ii z$ in \eqref{eq:Bessel},
\begin{align}
\label{eq:mod_Bessel}
\Big(\frac{\dd^2}{\dd z^2} + \frac{1}{z} \frac{\dd }{\dd z}
- 1 - \frac{\nu^2}{z^2}\Big) f(z)=0
\end{align}
has the standard solutions \cite{NIST}
\begin{align}
\label{eq:def_I}
I_\nu(z) =
\sum_{k=0}^\infty \frac{\big(\tfrac{z}{2}\big)^{2k+\nu}}{k!\Gamma(1+\nu+k)}
\end{align}
with definite behavior $\sim\big(\frac{z}{2}\big)^\nu$ at $z=0$ if
$\nu\notin\{-1,-2,\dots\}$ and $K_\nu(z)$ with definite decay behavior
\begin{align}
\label{eq:K_asymp}
K_\nu(z) \sim \sqrt{\frac{\pi}{2 z}} \ee^{-z}\word{as}
z\to \infty,\quad |\arg(z)|\leq \frac{3\pi}{2}-\delta.
\end{align}

$K_\nu$ and $I_\nu$ are related via the connection formula
\begin{align}
\label{eq:connection_KI}
K_\nu(z)=K_{-\nu}(z) = \frac{\pi}{2} \frac{I_{-\nu}(z)-I_\nu(z)}{\sin(\pi\nu)}.
\end{align}
The modified Bessel functions are related to the Bessel functions
by the following connection formulas (valid for $-\pi \leq \pm \arg(z) \leq \tfrac{\pi}{2}$):
\begin{align}
K_\nu(z) &= \pm\frac{\pi\ii}{2}\ee^{\pm\ii\frac{\pi\nu}{2}}
H^{(\pm)}_\nu\big( \ee^{\pm\ii\frac{\pi}{2}} z \big),\quad
\label{eq:connect_KH} \\ \label{eq:connect_IJ}
I_\nu(z) &= \ee^{\mp\ii\frac{\pi\nu}{2}} J_\nu\big(\ee^{\pm\ii\frac{\pi}{2}}z\big).
\end{align}

\subsection{Gegenbauer functions}
The propagators that we are able to compute explicitly are given in
terms of \emph{Gegenbauer functions}, which are solutions of the
\emph{Gegenbauer equation}
\begin{align}
\label{eq:Gegenbauer}
\Big((1-w^2) \frac{\dd^2}{\dd w^2} -2(1+\mu) w \frac{\dd }{\dd w} +\nu^2
- \big(\mu+\tfrac{1}{2}\big)^2\Big) f(w)=0.
\end{align}
The Gegenbauer equation is equivalent to the better-known associated
Legendre equation. In our context, however, it is easier to use
Gegenbauer functions instead of associated Legendre functions. For
details on Gegenbauer functions and their relation to associated 
Legendre functions, see \cite{DGR23a}.

The standard solutions of \eqref{eq:Gegenbauer} can be expressed in terms
of Gauß's hypergeometric function
\begin{subequations}
\begin{align}
 {\bf Z}_{\mu,\nu}(w)
 :=&\; \frac{(w\pm1)^{-\frac12-\mu-\nu}}{\Gamma(1+\nu)}
 {_2F_1}\Big(\tfrac12+\nu, \tfrac12+\mu+\nu;
 1+2\nu; \tfrac{2}{1\pm w}\Big)
 \notag \\ \label{eq:def_Z}
 =&\; \frac{w^{-\frac12-\mu-\nu}}{\Gamma(1+\nu)}
 {_2F_1}\Big(\tfrac12\big(\tfrac12+\nu+\mu\big),
 \tfrac12\big( \tfrac32+\mu+\nu\big);
 1+\nu; \tfrac{1}{w^2}\Big),
 \\ \label{eq:def_S}
 {\bf S}_{\mu,\nu}(w)
:=&\; \frac{1}{\Gamma(1+\mu)}
{_2F_1}\Big(\tfrac12+\nu+\mu, \tfrac12-\nu+\mu;
 1+\mu; \tfrac{1-w}{2}\Big).
\end{align}
\end{subequations}
The function ${\bf Z}_{\mu,\nu}(w)$ is holomorphic on $\bC\setminus]-\infty,1]$
and has definite behavior as $w\to\infty$:
\begin{align}
\label{eq:Z_asymp}
\frac{w^{-\frac12-\mu-\nu}}{\Gamma(1+\nu)} \word{as} w\to\infty.
\end{align}
The function ${\bf S}_{\mu,\nu}(w)$ is holomorphic on the larger set
$\bC\setminus]-\infty,-1]$ and is characterized by its regular behavior
at $w=1$, where we have ${\bf S}_{\mu,\nu}(1)=(\Gamma(1+\nu))^{-1}$.

Near $w=1$, we have for $\Re(\mu)>0$ \cite{DGR23a}:
\begin{align}
\label{eq:Z_at_pole}
{\bf Z}_{\mu,\nu}(w) &\sim
\frac{2^{-\frac12+\nu}\Gamma(\mu)}{\sqrt{\pi} \Gamma\big(\tfrac12+\mu+\nu\big) (w-1)^\mu},
 \notag \\
{\bf S}_{\mu,\nu}(-w) &\sim
\frac{2^{\mu}\Gamma(\mu)}{\Gamma\big(\tfrac12+\mu+\nu\big)
\Gamma\big(\tfrac12+\mu-\nu\big) (1-w)^{\mu}}.
\end{align}

\subsection{Integral identities}
An important integral identity involving Bessel functions that we need is
the following special case ($\alpha=\beta$) of the \emph{discontinuous
Weber-Schafheitlin integral} \cite{GR,Erdelyi2}:
\begin{align}
&\;\int_0^\infty J_{\nu}(\alpha z) J_{\mu}(\alpha z) z^{-\lambda} \dd z
\notag \\ \label{eq:dWS_v1}
=&\; \frac{\alpha^{\lambda-1}\Gamma(\lambda)\Gamma\big(\tfrac{\nu+\mu-\lambda+1}{2}\big)
}{2^\lambda \Gamma\big(\tfrac{-\nu+\mu+\lambda+1}{2}\big)
\Gamma\big(\tfrac{\nu+\mu+\lambda+1}{2}\big)
\Gamma\big(\tfrac{\nu-\mu+\lambda+1}{2}\big) }, \quad
\Re(\mu+\nu+1)>\Re(\lambda)>0,\quad \alpha>0.
\end{align}
In the case $\lambda=1$, we may also set $\alpha=1$ and obtain
for $\Re(\mu+\nu)>0$:
\begin{align} \label{eq:dWS_int}
\int_0^\infty J_{\nu}( z) J_{\mu}(z) \frac{\dd z}{z}
&= \frac{1}{(\mu+\nu) \Gamma\big(1-\tfrac{\nu-\mu}{2}\big)
\Gamma\big(1+\tfrac{\nu-\mu}{2}\big) }
= \frac{\sin\big(\pi \tfrac{\nu-\mu}{2} \big)}{(\mu+\nu) \pi \tfrac{\nu-\mu}{2}}.
\end{align}
Using the connection formula \eqref{eq:connect_HJ} and some trivial
transformations, we obtain, provided that
simultaneously $\Re(\mu+\nu)>0$ and $\Re(\mu-\nu)>0$,
\begin{align} \label{eq:dWS_HJ}
\int_0^\infty H^{(\pm)}_{\nu}( z) J_{\mu}(z) \frac{\dd z}{z}
%&= \frac{\pm\ii}{\sin(\pi\nu)}
%\int_0^\infty \Big( \ee^{\mp\ii\pi\nu} J_\nu(z)
%  - J_{-\nu}(z)\Big) J_{\mu}(z) \frac{\dd z}{z}
%&= \frac{\pm\ii}{\sin(\pi\nu)}
%\Big( \ee^{\mp\ii\pi\nu}
%\frac{\sin\big(\pi \tfrac{\nu-\mu}{2} \big)}{(\mu+\nu) \pi \tfrac{\nu-\mu}{2}}
%  + \frac{\sin\big(\pi \tfrac{\nu+\mu}{2} \big)}{(\nu-\mu) \pi \tfrac{\nu+\mu}{2}}\Big)
%&= \frac{\pm2\ii}{\pi \sin(\pi\nu) (\nu^2-\mu^2)}
%\Big( \ee^{\mp\ii\pi\nu} \sin\big(\pi \tfrac{\nu-\mu}{2} \big)
% + \sin\big(\pi \tfrac{\nu+\mu}{2} \big) \Big)
&= \frac{\pm2\ii \ee^{\pm\ii\pi\tfrac{\mu-\nu}{2}}}{\pi  (\nu^2-\mu^2)}.
\end{align}

To compute the spacetime representation of the propagators from the
resolvents of the one-dimensional Schrödinger operators with exponential
potentials, we need the following integrals. By \cite[6.692]{GR}, we have for
$\Re(a)>|\Re(b)|$, $c>0$ and $\Re(\nu)>-\frac32$:
\begin{align}
 \int_0^\infty x K_\nu(ax) I_\nu(bx)\sin(cx) \dd x
 &= -\frac{c}{2(ab)^{\frac32}}
  (w^2-1)^{-\frac12} Q_{\nu-\frac12}^1(w)
  \notag \\
  &=  \frac{\sqrt{\pi}\Gamma\big(\tfrac{3}{2}+\nu\big)}{2^{\nu}} \frac{c}{(2ab)^{\frac32}}
 {\bf Z}_{1,\nu}(w),
 \notag \\ \label{eq:KIsin_int}
 \word{where} w &= \frac{a^2+b^2+c^2}{2ab},
\end{align}
and where $Q^{\mu}_\nu$ is the associated Legendre
function of the second kind (see e.g. \cite{NIST})
and ${\bf Z}_{\mu,\nu}$ is the Gegenbauer function
\eqref{eq:def_Z}. If $\nu=\ii\lambda$, as is the case for 
our application, the integral \eqref{eq:KIsin_int} is a 
special case of a Kontorovich-Lebedev transform \cite{Oberhettinger, 
PathakPandey}.

\section{The Schrödinger operator with negative exponential potential}
\label{app:neg_exp}
In this appendix, we describe the Schrödinger operator with negative 
exponential potential. The boundary conditions and resolvents 
corresponding to all closed realizations of this operator, 
including the self-adjoint ones, have recently been described  
\cite{DeLee,Stempak}. For the convenience of the reader, we summarize these 
results in Subsection \ref{app:closed_sa_real}. In Subsection 
\ref{app:mode_rep}, we determine the basis of (generalized) eigenfunctions 
corresponding to the self-adjoint realizations and describe 
the resolvents as mode integrals/sum over these eigenfunctions. 
In Subsection \ref{app:ortho}, we verify the orthogonality of 
all eigenfunctions.
 
\subsection{Resolvent for closed and self-adjoint realizations}
\label{app:closed_sa_real}
Consider for $l>0$ the formal expression
\begin{align}
 M_{\ii l} := -\del_\eta^2 - l^2 \ee^{2\eta} \word{on} L^2(\bR).
\end{align}
Let us define $\cD_l := \{ f\in L^2(\bR) \;|\; M_{\ii l}f\in L^2(\bR) \}$.
There is a one-parameter family of closed realizations of $M_{\ii l}$
parametrized by $\gamma\in\bC\cup\{\infty\}$ \cite{DeLee}. Their
domains are given by
\begin{align}
 \cD\big( M_{\ii l}^\gamma\big) =
 \begin{cases}
  \Big\{ f\in \cD_l \;|\; \lim_{\eta\to\infty} \mathcal{W}\Big(
  H_{\frac12}^{(+)}(l \ee^\eta)+\gamma  H_{\frac12}^{(-)}(l \ee^\eta),f(\eta)\Big)=0 \Big\},
  &\; \gamma\in\bC,\vspace{11pt} \\
  \Big\{ f\in \cD_l \;|\; \lim_{\eta\to\infty} \mathcal{W}\Big(
   H_{\frac12}^{(-)}(l \ee^\eta),f(\eta)\Big)=0 \Big\},
  &\; \gamma=\infty,
 \end{cases}
\end{align}
corresponding to specific boundary conditions at $\eta\to\infty$. Here,
$\mathcal{W}(f,g) := fg'-f'g$ denotes the Wronskian and $H^{(\pm)}_\nu(z)$
are the Hankel functions \eqref{eq:def_Hankel}. By \eqref{eq:asymp_H}, we have
\begin{align}
H_{\frac12}^{(\pm)}(z) \sim \mp\ii\sqrt{\frac{2}{\pi z}}
\ee^{\pm\ii z },
\word{as} z\to\infty.
\end{align}

$ M_{\ii l}^0$ and
$M_{\ii l}^\infty$ are distinguished because they are the only closed
realizations that have purely continuous spectrum
\begin{align}
 \sigma(M_{\ii l}^0) =\sigma( M_{\ii l}^\infty) = [0,\infty[.
\end{align}

Now let $\gamma=\ee^{\ii\pi\alpha}\in\bC\setminus\{0\}$, then\footnote{
Note that the point spectrum has been computed incorrectly in an early
preprint of \cite{DeLee}, as pointed out by one of us.
}
\begin{align}
 \sigma(M_{\ii l}^\gamma) &= [0,\infty[ \; \cup \; \{-(2n+\alpha)^2  \;|\;
 \Re(\alpha+2n)>0,\; n\in\bZ \}, \notag \\
 \gamma&=\ee^{\ii\pi\alpha}\in\bC\setminus\{0\}.
\end{align}
The point spectrum is only real if $\alpha$ is real, $\alpha\in[0,2[$, or
if $|\gamma|=1$, respectively. In this case, the realization is self-adjoint.
Away from the spectrum of $M_{\ii l}^\gamma$, the integral kernel
$R_{\ii l}^\gamma(-\nu^2;\eta,{\eta'})$ of the resolvent 
$(M_{\ii l}^\gamma+\nu^2)^{-1}$ is given by \cite{DeLee}
\begin{align}
 R_{\ii l}^\gamma(-\nu^2;\eta,{\eta'}) = \frac{\pi\ii}{2(\ee^{\ii\pi \nu}-\gamma)}
 &\Big( \theta({\eta'}-\eta) J_{\nu}(l\ee^\eta) \big( \ee^{\ii\pi\nu} H^{(+)}_\nu(l \ee^{\eta'})
 + \gamma H^{(-)}_\nu(l\ee^{\eta'}) \big)
 \notag \\ \label{eq:resol_neg_exp}
 &+\theta(\eta-{\eta'}) J_{\nu}(l\ee^{\eta'}) \big( \ee^{\ii\pi\nu} H^{(+)}_\nu(l \ee^\eta)
 + \gamma H^{(-)}_\nu(l\ee^\eta) \big)\Big),
\end{align}
where $J_\nu(z)$ is the Bessel function \eqref{eq:defJ}.

\subsection{Basis of (generalized) eigenfunctions and 
integral representation of the resolvent}
\label{app:mode_rep}
Let now $\gamma=\ee^{\ii\pi\alpha}$ with $\alpha\in[0,2[$. For these values
of $\gamma$, corresponding to self-adjoint realizations, Stone's
formula allows us to obtain the spectral projections and hence the 
mode functions from the resolvent and
write the resolvent as a mode integral. Since the spectrum has a continuous
and a discrete part, this integral will also have a continuous and a discrete part.
We will show in the following that for $-\nu^2\notin\sigma\big(M_{\ii l}^\gamma\big)$,
we have
\begin{align}
\label{eq:resolvent_sa_mode_sum}
R_{\ii l}^\gamma(-\nu^2;\eta,\eta')
= \int_0^\infty \frac{f_{\gamma,\lambda,l}(\eta) f_{\gamma,\lambda,l}^\ast(\eta')
}{\lambda^2+\nu^2}
2\lambda\dd\lambda
+\sum_{n=0}^\infty  \frac{g_{\gamma,n,l}(\eta)g_{\gamma,n,l}^\ast(\eta')
}{-(\alpha+2n)^2+\nu^2},
\end{align}
where
\begin{align} \label{eq:fgala}
f_{\gamma,\lambda,l}(\eta)
 &= \frac{\sqrt{\sinh(\pi\lambda)}}{2|\ee^{\ii\pi\alpha}-\ee^{-\pi\lambda}|}
 \Big(H^{(-)}_{\ii\lambda}(l\ee^\eta)
 + \ee^{-\pi(\lambda+\ii\alpha)}  H^{(+)}_{\ii\lambda}(l\ee^\eta)  \Big),
 \\ \label{eq:ggala}
 g_{\gamma,n,l}(\eta) &= \sqrt{2(\alpha+2n)} J_{\alpha+2n}(l\ee^\eta),\quad
 n=\begin{cases}
 0,1,2,\dots,&\quad\alpha > 0, \\
 1,2,3,\dots,&\quad \alpha=0,
 \end{cases}
\end{align}
and where we made the slight abuse of notation $g_{1,0,l}(\eta)\equiv0$ for $\alpha=n=0$
in \eqref{eq:resolvent_sa_mode_sum}. We separately derive the mode functions
corresponding to the continuous spectrum and to the discrete spectrum.

\paragraph{Continuous spectrum.}
Let $\lambda>0$. We find for ${\eta'}>\eta$ that
\begin{align}
 &R_{\ii l}^\gamma(\lambda^2+\ii0;\eta,{\eta'})- R_{\ii l}^\gamma(\lambda^2-\ii0;\eta,{\eta'})
 \notag \\
 =&\; R_{\ii l}^\gamma\big(-(-\ii \lambda+0)^2;\eta,{\eta'}\big)
 - R_{\ii l}^\gamma\big(-(\ii \lambda+0)^2;\eta,{\eta'}\big)
 \notag \\
 =&\; \frac{\pi\ii}{2} \Bigg( \frac{J_{-\ii\lambda}(l \ee^\eta)}{\ee^{\pi\lambda}-\gamma}
 \Big( \ee^{\pi\lambda} H^{(+)}_{-\ii\lambda}(l\ee^{\eta'})
 + \gamma H^{(-)}_{-\ii\lambda}(l\ee^{\eta'})\Big)
 - (\lambda\leftrightarrow-\lambda)\Bigg).
\end{align}
The connection formula \eqref{eq:connect_H_pmnu} yields
\begin{align}
 \ee^{\pi\lambda} H^{(+)}_{-\ii\lambda}(l\ee^{\eta'})
 + \gamma H^{(-)}_{-\ii\lambda}(l\ee^{\eta'})
 &=  H^{(+)}_{\ii\lambda}(l\ee^{\eta'})
 + \gamma \ee^{\pi\lambda} H^{(-)}_{\ii\lambda}(l\ee^{\eta'}),
\end{align}
hence for ${\eta'}>\eta$
\begin{align}
 &R_{\ii l}^\gamma(\lambda^2+\ii0;\eta,{\eta'})- R_{\ii l}^\gamma(\lambda^2-\ii0;\eta,{\eta'})
 %\notag \\
 %=&\; \frac{\pi\ii}{2} \Bigg( \frac{\ee^{\pi\lambda}}{\ee^{\pi\lambda}-\gamma}J_{-\ii\lambda}(l \ee^\eta)
 %- \frac{1}{\ee^{-\pi\lambda}-\gamma}J_{\ii\lambda}(l \ee^\eta)\Bigg)
 % \Big( \ee^{-\pi\lambda} H^{(+)}_{\ii\lambda}(l\ee^{\eta'})
 %+ \gamma H^{(-)}_{\ii\lambda}(l\ee^{\eta'})\Big)
 \notag \\
 =&\; \frac{\pi\ii \Big(  (1-\ee^{\pi\lambda} \gamma) J_{-\ii\lambda}(l \ee^\eta)
 -(\ee^{\pi\lambda}-\gamma) J_{\ii\lambda}(l \ee^\eta)\Big)
  \Big( \ee^{-\pi\lambda} H^{(+)}_{\ii\lambda}(l\ee^{\eta'})
 + \gamma H^{(-)}_{\ii\lambda}(l\ee^{\eta'})\Big)}{2(\ee^{\pi\lambda}-\gamma)(\ee^{-\pi\lambda}-\gamma)}.
\end{align}
Now the connection formulas \eqref{eq:connect_JH} and \eqref{eq:connect_H_pmnu} imply
\begin{align}
 (1-\ee^{\pi\lambda} \gamma) J_{-\ii\lambda}
 -(\ee^{\pi\lambda}-\gamma) J_{\ii\lambda}
 %\notag \\
 %=&\;\frac12 \Big(
 %(1-\ee^{\pi\lambda} \gamma) \big( \ee^{-\pi\lambda}H^{(+)}_{\ii\lambda}
 %+\ee^{\pi\lambda}H^{(-)}_{\ii\lambda}\big)
 %-(\ee^{\pi\lambda}-\gamma) \big(H^{(+)}_{\ii\lambda}+H^{(-)}_{\ii\lambda}\big)
 %\Big)
 %\notag \\
 =&\; - \sinh(\pi\lambda)\ee^{\pi\lambda}\gamma \Big(H^{(-)}_{\ii\lambda}
 + \ee^{-\pi\lambda} \gamma^\ast H^{(+)}_{\ii\lambda}  \Big).
\end{align}
Furthermore,  because for $z>0$ we have $H^{(\pm)\ast}_{\ii\lambda}(z)
= \ee^{\pm\pi\lambda}H^{(\mp)}_{\ii\lambda}(z)$, we find
\begin{align}
\ee^{-\pi\lambda} H^{(+)}_{\ii\lambda}(l\ee^{\eta'})
 + \gamma H^{(-)}_{\ii\lambda}(l\ee^{\eta'})
=\Big( H^{(-)}_{\ii\lambda}(l\ee^{\eta'})
 + \ee^{-\pi\lambda} \gamma^\ast H^{(+)}_{\ii\lambda}(l\ee^{\eta'})\Big)^\ast.
\end{align}
Using the symmetry of the following expression under complex conjugation, we
finally obtain for all $\eta$ and ${\eta'}$:
\begin{align}
 &\frac{1}{2\pi\ii}\Big(R_{\ii l}^\gamma(\lambda^2+\ii0;\eta,{\eta'})
 - R_{\ii l}^\gamma(\lambda^2-\ii0;\eta,{\eta'})\Big)
 \notag \\
 =&\; \frac{\sinh(\pi\lambda)  }{4|\ee^{\ii\pi\alpha}-\ee^{-\pi\lambda}|^2}
\Big(H^{(-)}_{\ii\lambda}(l\ee^\eta)
 + \ee^{-\pi(\lambda+\ii\alpha)}  H^{(+)}_{\ii\lambda}(l\ee^\eta)  \Big)
 \Big(H^{(-)}_{\ii\lambda}(l\ee^{\eta'})
 + \ee^{-\pi(\lambda+\ii\alpha)}  H^{(+)}_{\ii\lambda}(l\ee^{\eta'})  \Big)^\ast .
\end{align}
This yields the mode functions \eqref{eq:fgala}.

\paragraph{Point spectrum.}
Let now $\nu=\alpha+2n$ for some $n\in\bN_0$. Then using the connection
formula \eqref{eq:connect_JH}, we find
\begin{align}
 &\quad-\ii \lim_{\eps\downarrow0} \eps R_{\ii l}^\gamma(-(\alpha+2n)^2+\ii\eps;\eta,{\eta'})
 \notag \\
 &=  \pi J_{\alpha+2n}(l\ee^\eta) J_{\alpha+2n}(l\ee^{\eta'})\lim_{\eps\downarrow0}
 \frac{\eps}{\ee^{\pi\frac{\eps}{2(\alpha+2n)}}-1}
 \notag \\
 &= 2(\alpha+2n) J_{\alpha+2n}(l\ee^\eta) J_{\alpha+2n}(l\ee^{\eta'}).
\end{align}
Thus, the mode functions corresponding to the point spectrum are given by
\eqref{eq:ggala}.

\subsection{Orthogonality of the mode functions}
\label{app:ortho}
Let us reintroduce the timelike variable $\tau=R\ee^\eta$. As (generalized)
eigenfunctions of a self-adjoint operator, the mode functions $f_{\gamma,n,l}(\tau)$
and $g_{\gamma,n,l}(\tau)$ defined in \eqref{eq:fgala} and \eqref{eq:ggala}
are orthogonal and complete for fixed $\gamma=\ee^{\ii\pi\alpha}$
with $\alpha\in[0,2[$. In this section, we
verify orthogonality by hand.

The orthogonality relations of the mode functions corresponding to the point
spectrum are a special case of the \emph{discontinuous Weber-Schafheitlin integral}
\eqref{eq:dWS_int}. Thus,
\begin{align}
\int_0^\infty g_{\gamma,n,l}(\tau) g_{\gamma,n',l}(\tau) \frac{\dd \tau}{\tau}
= \frac{\sqrt{(\alpha+2n)(\alpha+2n')}}{(\alpha+n+n')}
 \frac{\sin\big(\pi(n-n')\big)}{\pi(n-n')} = \delta_{n,n'}.
\end{align}

The orthogonality of the mode functions corresponding to the point spectrum
to all mode functions corresponding to the continuous spectrum follows from
the integral \eqref{eq:dWS_HJ}, which is in turn obtained from \eqref{eq:dWS_int}
together with the connection formula \eqref{eq:connect_HJ}. We may also use the
reality of $g_{\gamma,n,l}(\tau)$ to compute the scalar product:
\begin{align}
&\;\int_0^\infty f_{\gamma,\lambda,l}(\tau) g_{\gamma,n,l}(\tau) \frac{\dd \tau}{\tau}
\notag \\
=&\; c_{\alpha,\lambda}
\int_0^\infty \Big(H^{(-)}_{\ii\lambda}(\tau)
 + \ee^{-\pi(\lambda+\ii\alpha)}  H^{(+)}_{\ii\lambda}(\tau)  \Big) J_{\alpha+2n}(\tau) \frac{\dd \tau}{\tau}
\\ \notag
=&\;0,
\end{align}
 where $c_{\alpha,\lambda}=\frac{\sqrt{2(\alpha+2n)\sinh(\pi\lambda)}}{2 |\ee^{\ii\pi\alpha}-\ee^{-\pi\lambda}|}$.

Most intricate is the orthogonality of $f_{\gamma,\lambda,l}$ and $f_{\gamma,\lambda',l}$.
We first use the connection formula \eqref{eq:connect_HJ} to 
express $f_{\gamma,\lambda,l}(\tau)$ solely in terms of the Bessel functions 
$J_{\ii\lambda}(l\tau)$ and $J_{-\ii\lambda}(l\tau)$.
% \begin{align}
% f_{\gamma,\lambda,l}(\tau)
% = \frac{J_{\ii\lambda}(l\tau) \big(\ee^{-\ii\pi\alpha}-\ee^{-\pi\lambda}\big)
% +J_{-\ii\lambda}(l\tau) \big(1-\ee^{-\pi\lambda}\ee^{-\ii\pi\alpha}\big)
% }{2\sqrt{\sinh(\pi\lambda)}  |\ee^{\ii\pi\alpha}-\ee^{-\pi\lambda}|}.
% \end{align}
This yields the inner product as the unique distributional boundary value
\begin{align}
&\;4\sqrt{\sinh(\pi\lambda)\sinh(\pi\lambda')}  |\ee^{\ii\pi\alpha}-\ee^{-\pi\lambda}|
 |\ee^{\ii\pi\alpha}-\ee^{-\pi\lambda'}|
\int_0^\infty f_{\gamma,\lambda,l}^\ast(\tau) f_{\gamma,\lambda',l}(\tau) \frac{\dd \tau}{\tau}
\notag \\
=&\; \lim_{\eps\searrow0} \int_0^\infty
\Big(
J_{-\ii(\lambda+\ii\eps)}(\tau) J_{\ii\lambda'}(\tau)\big(\ee^{\ii\pi\alpha}-\ee^{-\pi\lambda}\big)
\big(\ee^{-\ii\pi\alpha}-\ee^{-\pi\lambda'}\big)
\notag \\
&\hspace{9ex} + J_{\ii(\lambda-\ii\eps)}(\tau) J_{\ii\lambda'}(\tau)
\big(1-\ee^{-\pi\lambda}\ee^{\ii\pi\alpha}\big)
\big(\ee^{-\ii\pi\alpha}-\ee^{-\pi\lambda'}\big)
\notag \\
&\hspace{9ex} + J_{-\ii(\lambda+\ii\eps)}(\tau) J_{-\ii\lambda'}(\tau)
\big(\ee^{\ii\pi\alpha}-\ee^{-\pi\lambda}\big)
\big(1-\ee^{-\pi\lambda'}\ee^{-\ii\pi\alpha}\big)
\notag \\
&\hspace{9ex} + J_{\ii(\lambda-\ii\eps)}(\tau) J_{-\ii\lambda'}(\tau)
\big(1-\ee^{-\pi\lambda}\ee^{\ii\pi\alpha}\big)
\big(1-\ee^{-\pi\lambda'}\ee^{-\ii\pi\alpha}\big)
\Big) \frac{\dd \tau}{\tau},
\end{align}
which can be analyzed by the Weber-Schafheitlin integral as before.
We obtain 
\begin{align}
&\;-2\pi\ii\sqrt{\sinh(\pi\lambda)\sinh(\pi\lambda')}  |\ee^{\ii\pi\alpha}-\ee^{-\pi\lambda}|
 |\ee^{\ii\pi\alpha}-\ee^{-\pi\lambda'}|
\int_0^\infty f_{\gamma,\lambda,l}^\ast(\tau) f_{\gamma,\lambda',l}(\tau) \frac{\dd \tau}{\tau}
% \notag \\
% =&\; \frac{\ee^{\ii\pi\alpha}-\ee^{-\pi\lambda}
% }{\lambda^2-{\lambda'}^2+\ii0}
% \Big( \sinh\big(\pi\tfrac{\lambda+\lambda'}{2}\big)
% \big(\ee^{-\ii\pi\alpha}-\ee^{-\pi\lambda'}\big)
% +\sinh\big(\pi\tfrac{\lambda-\lambda'}{2}\big)
% \big(1-\ee^{-\pi\lambda'}\ee^{-\ii\pi\alpha}\big) \Big)
% \notag \\
% &- \frac{1-\ee^{\ii\pi\alpha}\ee^{-\pi\lambda}
% }{\lambda^2-{\lambda'}^2-\ii0}
% \Big( \sinh\big(\pi\tfrac{\lambda+\lambda'}{2}\big)
% \big(1-\ee^{-\pi\lambda'}\ee^{-\ii\pi\alpha}\big)
% +\sinh\big(\pi\tfrac{\lambda-\lambda'}{2}\big)
% \big(\ee^{-\ii\pi\alpha}-\ee^{-\pi\lambda'}\big) \Big)
\notag \\
=&\; \frac12
\Big( \ee^{\pi\tfrac{\lambda+\lambda'}{2}} - \ee^{-3\pi\tfrac{\lambda+\lambda'}{2}}
+\ee^{-\pi\tfrac{3\lambda-\lambda'}{2}}-\ee^{\pi\tfrac{\lambda-3\lambda'}{2}}
+2\cos(\pi\alpha) \big(\ee^{-\pi\tfrac{\lambda+3\lambda'}{2}}-
\ee^{-\pi\tfrac{\lambda-\lambda'}{2}}\big)
\Big) 
\notag \\
&\;\times
\Bigg( \frac{1}{\lambda^2-{\lambda'}^2+\ii0}- \frac{1}{\lambda^2-{\lambda'}^2-\ii0} \Bigg).
\end{align}
Then the Sokhotski–Plemelj formula
\begin{align}\label{eq:Sokho}
\frac{1}{x+\ii0}-\frac{1}{x-\ii0}=-2\pi\ii\delta(x)
\end{align}
together with $\lambda,\lambda'\geq0$ imply
\begin{align}
\int_0^\infty f_{\gamma,\lambda,l}^\ast(\tau) f_{\gamma,\lambda',l}(\tau) \frac{\dd \tau}{\tau}
= \delta(\lambda^2-{\lambda'}^2),
\end{align}
which is the last missing orthogonality relation.

\paragraph{Acknowledgment.}
CG would like to thank Jan Dereziński for helpful discussions regarding the
Bessel operator and Schrödinger operators with exponential potentials.
HS would like to thank Pei-Ming Ho and
Hikaru Kawai for discussions in a related collaboration, as well as Masanori Hanada for early discussions on the $k=0$ spacetime.
This work is supported by the Austrian Science Fund (FWF) grant P36479.

\end{document}